\documentclass[useAMS,usenatbib]{mn2e_mod}
\usepackage{natbib}
\usepackage{graphicx}
\usepackage{amsmath}
\usepackage{amssymb}
\usepackage{deluxetable}
\usepackage{longtable}
\usepackage{url}

\let\ACMmaketitle=\maketitle
\renewcommand{\maketitle}{\begingroup\let\footnote=\thanks \ACMmaketitle\endgroup}

\title[Black hole masses of type~II AGN]{Black hole mass estimation for Active Galactic Nuclei from a new angle}

\author[Baron \& M$\mathrm{\acute{e}}$nard]
{Dalya Baron$^{1}$\thanks{dalyabaron@gmail.com},
Brice M$\mathrm{\acute{e}}$nard$^{2}$
\\
$^{1}$School of Physics and Astronomy, Tel-Aviv University, Tel Aviv 69978, Israel.\\
$^{2}$Department of Physics and Astronomy, Johns Hopkins University, Baltimore, MD, 21218, USA\\
}

\begin{document}

\maketitle

\label{firstpage}
\begin{abstract}

The scaling relations between supermassive black holes and their host galaxy properties are of fundamental importance in the context black hole-host galaxy co-evolution throughout cosmic time. In this work, we use a novel algorithm that identifies smooth trends in complex datasets and apply it to a sample of $2\,000$ type~I active galactic nuclei (AGN) spectra. We detect a sequence in emission line shapes and strengths which reveals a correlation between the narrow L([OIII])/L(H$\beta$) line ratio and the width of the broad H$\alpha$. This scaling relation ties the kinematics of the gas clouds in the broad line region to the ionisation state of the narrow line region, connecting the properties of gas clouds kiloparsecs away from the black hole to material gravitationally bound to it on sub-parsec scales. This relation can be used to estimate black hole masses from narrow emission lines only. It therefore enables black hole mass estimation for obscured type~II AGN and allows us to explore the connection between black holes and host galaxy properties for thousands of objects, well beyond the local Universe. Using this technique, we present the $M_{\mathrm{BH}} - \sigma$ and $M_{\mathrm{BH}} - M_{*}$ scaling relations for a sample of about 10\,000 type~II AGN from SDSS. These relations are remarkably consistent with those observed for type~I AGN, suggesting that this new method may perform as reliably as the classical estimate used in non-obscured type~I AGN. These findings open a new window for studies of black hole-host galaxy co-evolution throughout cosmic time.

\end{abstract}


\vspace{1cm}
\section{Introduction}\label{s:intro}

A fascinating discovery of modern-day astronomy is that most galaxies host supermassive black holes in their centre. With this came a surprise: the mass of these black holes appears to be linked to the stellar velocity dispersion in the bulge, the bulge mass and bulge luminosity (e.g., \citealt{kormendy95, magorrian98, gebhardt00a, ferrarese00, tremaine02, marconi03, haring04, gultekin09}). It suggests that black holes may play a fundamental role in galaxy evolution but the exact details are still unclear (e.g. \citealt{silk98,kauffmann00}). These observational connections were established in the local universe in massive galaxies with dormant black holes, where black hole masses were measured through direct dynamical methods (e.g., \citealt{gebhardt00a,ferrarese96, marconi01,herrnstein05, kuo11}). So far, roughly a hundred dynamical black hole masses are available \citep{kormendy13,mcconnell13} with enough accuracy to probe black hole - host galaxy property relations.

Beyond the local universe, black hole masses are measured through indirect methods in active galactic nuclei (AGN). The unified model of AGN \citep{antonucci93, urry95, netzer15} consists of an accretion disk, fast-moving gas clouds in the vicinity of the black hole (the broad line region; BLR) surrounded by a dusty torus, and more slowly-moving gas residing on galactic scales that is photoionised by the central source (the narrow line region; NLR). The observed properties of AGN depend on the inclination of the system with respect to the observer. In unobscured type~I AGN, the line of sight probes the accretion disk, the BLR and the NLR. The radiation originating from the accretion disk outshines the stars, preventing the host galaxy to be observed. In obscured type~II AGN, the system is viewed edge-on with the dusty torus obscuring the bright accretion disk and the BLR but allowing us to see the NLR and the host galaxy. In such systems, black hole masses are estimated through an application of the virial method, where the BLR clouds are assumed to be following the gravitational potential of the black hole. The two main methods, reverberation mapping and single-epoch mass determination, require a direct view of the accretion disk and BLR clouds, and can thus only be applied in unobscured type~I AGN (e.g., \citealt{kaspi00, peterson04, greene05b, vestergaard06, kaspi05, bentz09, shen11, trakhtenbrot12, bentz13}). Despite the caveats and uncertainties in these methods (e.g., \citealt{vestergaard06, shen13}), measuring the black hole mass in type~I AGN is usually within reach. Unfortunately, for this type of objects the host galaxy properties are more difficult to obtain since the AGN dominates the observed flux in most wavelength ranges. Different techniques have been applied to separate the AGN from its host galaxy (e.g., \citealt{greene06_vel, peng06, woo08, merloni10, bennert15, reines15, shen15, bentz18}), each is subject to some uncertainties and limitations, in particular, most of the methods cannot be applied to the most luminous quasars, which completely outshine their hosts in all wavelengths. Studies have collected only a few hundreds of type~I AGN for which both the black hole mass and the host galaxy properties were measured. These samples also show a correlation between the black hole mass and its host properties, though their slopes differ, in some cases, from those found in passive galaxies in the local universe (e.g, \citealt{greene06, xiao11, reines15}), and they typically show a larger scatter around the best-fitting relations.

In this work, we make use of a novel algorithm we developed to find sequences in complex datasets. Applying it to a sample of about $2\,000$ type~I AGN spectra, we discover a new correlation that ties the ionisation state of the gas in the NLR to the kinematics of the clouds in the BLR (section \ref{s:sec_1}). This connection brings the missing piece required to estimate black hole masses in obscured type~II AGN. Using this new correlation, we estimate the black hole mass of about $10\,000$ type~II AGN and present the corresponding black hole - host galaxy scaling relations (section \ref{s:new_black_hole_mass_estimate}). We discuss our results in section \ref{s:disc}, and conclude in section \ref{s:concs}.

\section{A sequence of type~I AGN}\label{s:sec_1}

AGN spectra carry the signature of a variety of physical phenomena \citep{antonucci93,urry95,netzer15}: continuum and broad emission lines emitted by material in the vicinity of the black hole, narrow emission lines originating from gas on galactic scales illuminated by the accretion disk, and stellar light from the host galaxy. With a mixture of broad and narrow lines on top of each other and the stochasticity of accretion processes which affect the shape of the emission lines, the observed spectra display a significant level of complexity. Describing them quantitatively requires a non-trivial number of parameters and, as a result, the search for underlying trends between the different physical components has been challenging. To tackle this apparent complexity, one can use dimensionality reduction techniques. Principal component analysis (PCA) has been used in this context but most studies applied this technique on a set of measured parameters from the spectra rather than using the entire spectra. These chosen parameters are typically emission line widths and equivalent widths, line ratios, line asymmetry etc.  Using such a PCA analysis, \citet{boroson92} analysed a sample of 87 broad line AGN and revealed several correlations between their properties, in particular, they found that the main variance in these parameters is due to an anti-correlation between the strength of the broad FeII line and the narrow [OIII] line (see also \citealt{sulentic00, shen14}). While these results have provided valuable insight into the physics of AGN, it is important to search for other trends which might not necessarily show up in such parameter-based PCA approaches.

\begin{figure*}
\begin{center}
\includegraphics[width=.95\textwidth]{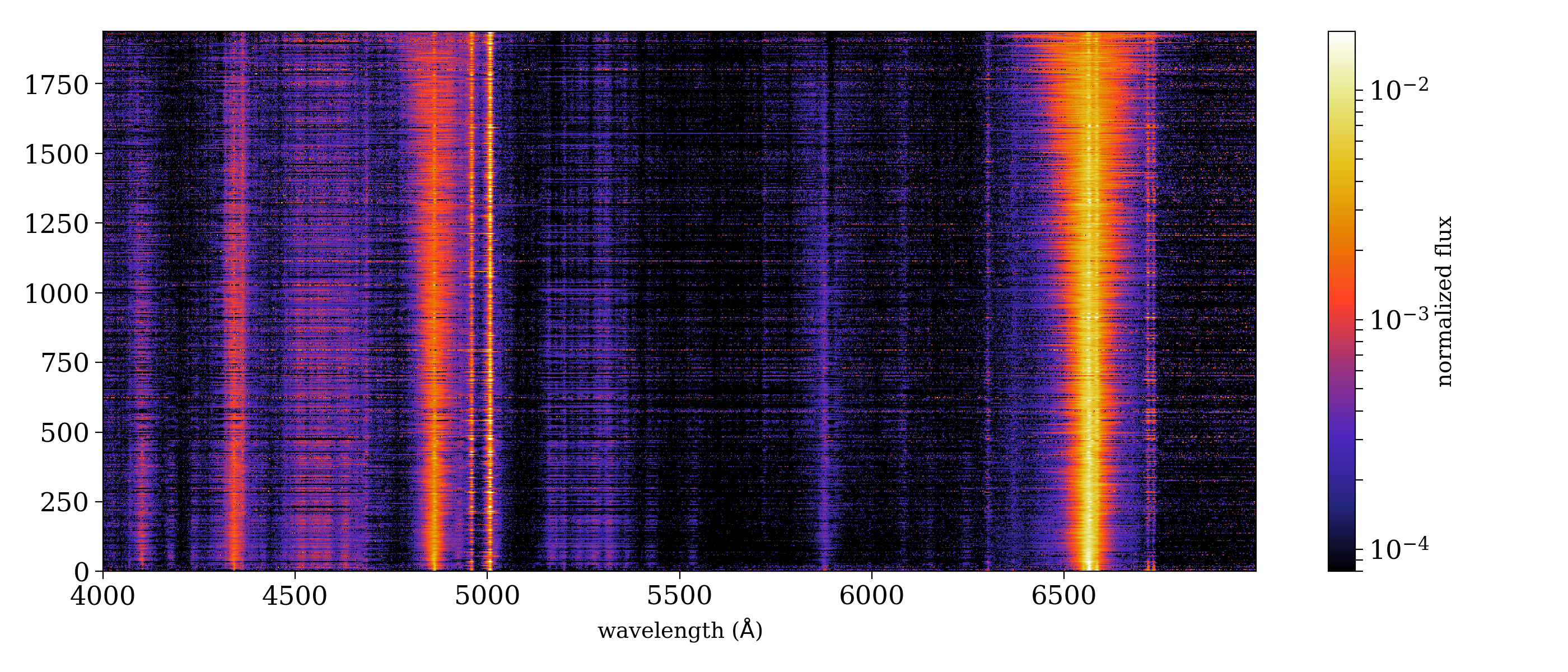}
\caption{The rest-frame AGN spectra ordered by line shape similarity. Each row represents an emission line spectrum colour-coded by normalised flux. The overall trend reveals a sequence primarily driven by the width and shape of the broad H$\alpha$ line.}
\label{f:sequence_result_spectra}
\end{center}
\end{figure*}

\subsection{The Sequencer}\label{s:sequencer_description}

We recently developed a new algorithm, the \emph{Sequencer} \citep{baron_menard}, which attempts 
to reveal the main sequence in a dataset, if it exists. To do so, it reorders objects within a set to produce the most elongated manifold describing their similarities which are measured in a multi-scale manner and using a collection of metrics. To be generic, it combines information from the Wasserstein distance \citep{rubner98}, the Kullback-Leibler divergence \citep{kullback51} and the L2 norm. This approach is pixel-based and allows us to consider the entire spectrum without the need to work with an arbitrary set of measured parameters. Contrarily to the traditional PCA approach, the Sequencer does not attempt to decompose the variation of the data onto a basis set. It only re-orders objects within a dataset, maximizing the similarities between neighbours and producing the most elongated manifold representing the entire dataset.
Such an approach can detect underlying trends not necessarily visible through PCA analyses. In appendix~\ref{ref:comparison_PCA}, we present a detailed comparison between the new sequence presented in this work with PCA-based results from the literature. 

The algorithm and its performance are presented in a separate paper. As described below, while this sequencing operation provides us with insight into the underlying trend present in the data, the main results of our analysis can be obtained without the use of this algorithm.

\subsection{The type~I AGN sample}\label{s:type1_sample}

Our goal is to explore the information content of AGN spectra, in particular the shape of some of the most important emission lines such as the hydrogen Balmer lines. To do so, we use spectroscopic data from the seventh data release (DR7) of the Sloan Digital Sky Survey (SDSS; \citealt{york00}). This dataset contains about a million of extragalactic spectra covering the wavelength range of 3800--9200\AA\ with 2.5\AA\ resolution. It includes 105\,783 type~I AGN brighter than $M_{i} = -22$ mag \citep{schneider10} with an average signal-to-noise ratio (SNR) of 10 pixel$^{-1}$. We select systems with $z<0.3$ so that the observed H$\alpha$ emission line stays at wavelengths shorter than 8500\AA, above which residual telluric lines become non-negligible. This reduces the sample size to 1941 AGN. We use redshift estimates based on the narrow lines as they are expected to be more accurate than those based on the broad lines \citep{hewett10}. \citet{shen11} present a catalogue of measured properties for these AGN which includes continuum and emission lines measurements around the H$\alpha$, H$\beta$, MgII, and CIV emission lines. These authors extracted emission line properties by fitting several kinematic components to account for the presence of broad and narrow emission lines. They used the former to estimate the virial black hole mass of each system \citep{vestergaard06}. 
    
In order to estimate the continuum flux, we shift the spectra to rest-frame wavelength and then interpolate them to a common wavelength grid from 4000\AA\ to 7000\AA\ with a 1\,\AA\ resolution. We select five spectral regions free of emission lines (4190--4215\AA, 5090-5110\AA, 5400--5750\AA, 6195--6215\AA, and 6840--6860\AA; \citealt{shen11,trakhtenbrot12,capellupo16}), estimate their flux distributions at a resolution of 30\AA, and use them as anchor points to perform a cubic spline interpolation. A visual inspection of the spectra confirms the validity of this approach. We repeated the process by varying the resolution from 15 to 50 \AA\ and obtained similar results. We subtract these continuum estimates from the initial spectra to obtain a set of optical emission line spectra.

\subsection{Sequence results}\label{s:sequence_results}
    
\begin{figure}
\includegraphics[width=0.5\textwidth]{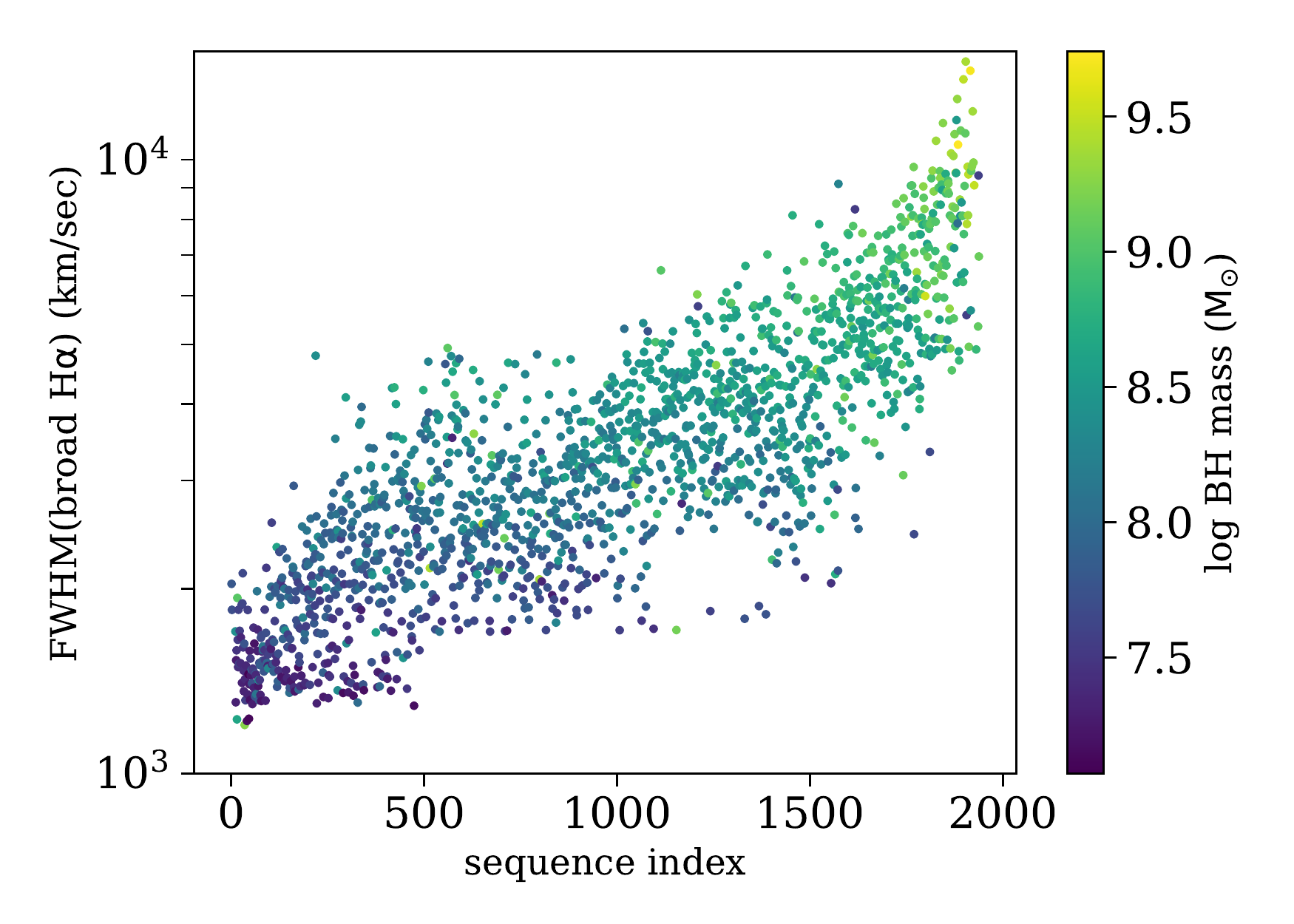}
\caption{The width of the broad H$\alpha$ line from the \citet{shen11} catalogue shown as a function of the sequence index of each spectrum. The distribution shows that the sequence is not only driven by the width of the Balmer line. Additional information originates from the shape of the emission lines. For reference, the data points are coloured according to the black hole mass of each object.}
\label{f:sequence_index_fwhm_bh_mass}
\end{figure}

We look for the existence of an underlying trend in the shape of the H$\alpha$ emission lines. To ensure that the broad H$\alpha$ line is fully observed, we focus on the wavelength range 6330--6800\,\AA\ which covers velocities of $\pm10,000$ km/sec with respect to the systemic velocity. We use the above algorithm with the selected sample of flux-normalised emission line spectra and obtain an ordering of the objects 
which is based on the generic similarity between neighbors. In Figure \ref{f:sequence_result_spectra} we show the corresponding flux-normalised emission line spectra. We observe an overall sequence which appears to be mainly driven by the width of the broad H$\alpha$ line. To explore this quantitatively, Figure \ref{f:sequence_index_fwhm_bh_mass} shows the FWHM of the broad H$\alpha$ as a function of the index of the objects along the sequence. In addition, the colours show the black hole mass estimates from the \citet{shen11} catalogue. The correlation between the FWHM of the broad H$\alpha$ and the sequence index shows a significant scatter. This indicates that the sequence is not only based on the width of the Balmer line but is informed by additional information which, in this case, is line shape. The broad emission lines in type~I AGN show complex, non-Gaussian, shapes and we indeed find that objects which are neighbours in the sequence have similar emission line profiles. In Figure \ref{f:correlation_with_sequence_index} we show the relation between the detected sequence and various measured properties of individual AGN, taken from \citet{shen11}.

\subsection{A relation between the Broad and Narrow Line Regions}\label{s:corr_NLR_BLR}
    
\begin{figure*}
\begin{center}
\includegraphics[width=\textwidth]{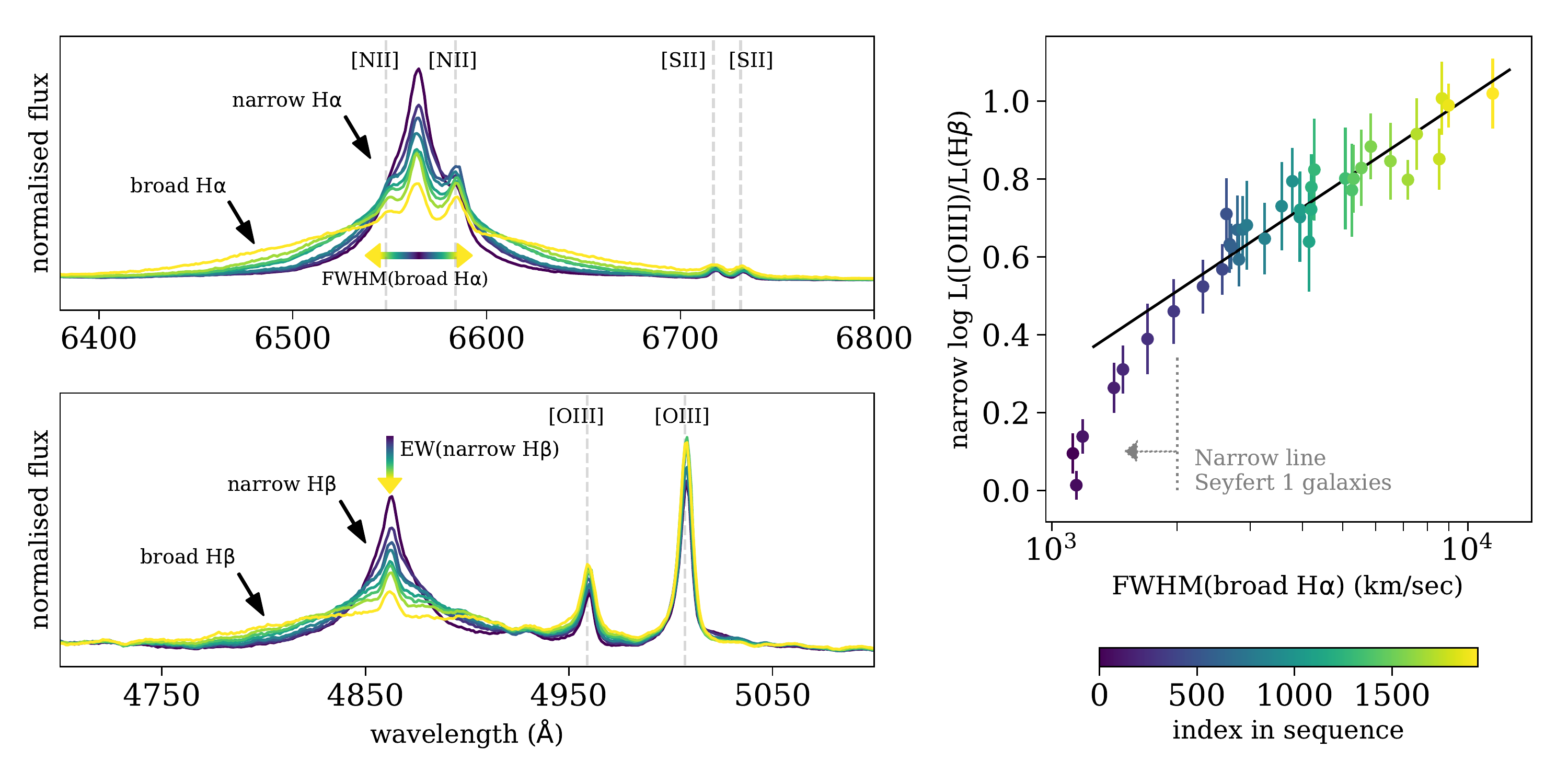}
\caption{\textbf{A sequence in the emission line shapes of unobscured AGN.} Left: eight median spectra along the sequence obtained by ordering about $2\,000$ AGN spectra based on multi-scale similarities. The two panels, showing the H$\alpha$ and H$\beta$ regions, reveal two trends: (1) the broad Balmer lines transition from wide Gaussian to narrower Lorentzian shapes while (2) the strength of the {narrow} H$\alpha$ and H$\beta$ lines decreases. Right panel: The narrow L([OIII])/L(H$\beta$) line ratio of median spectra as a function of the width of the broad H$\alpha$ for median spectra along the sequence. This correlation shows a connection between narrow and broad emission lines. It reveals the existence of a process connecting the broad and narrow line regions and shows that this narrow line ratio carries information about the kinematics of the gas in the vicinity of the black hole.}
\label{f:narrow_broad_lines_correlation}
\end{center}
\end{figure*}

To visually explore the dependence of narrow lines, we divide the sequence into eight consecutive bins. For each of them we estimate the median spectrum and show the results in the left panels of Figure \ref{f:narrow_broad_lines_correlation}. The top panel shows the H$\alpha$ and [NII] emission lines region and the bottom panel shows the H$\beta$ and [OIII] region. The stacked spectra show a clear trend throughout the sequence. First, the shape of the broad lines changes continuously with a transition from wider, Gaussian broad lines to narrower, Lorentzian lines. This well-known trend is a subject of extensive study, e.g., studies explore the relative contribution of turbulence versus virial rotation to the observed line shapes, and others examine the correlation of the shape with the AGN accretion rate and orientation (e.g., \citealt{sulentic02,collin06,marziani09,kollatschny13}). In addition, we observe that on average the flux of the narrow Balmer lines decreases as the broad lines become wider. This trend is not observed for the other narrow lines present in these two wavelength ranges, showing that the change in the narrow H$\alpha$ and H$\beta$ fluxes is not due to the overall flux normalisation. To further investigate this trend, we consider the narrow line ratio L([OIII])/L(H$\beta$). Being a ratio, it is independent of the overall flux normalisation and, being based on closely separated lines, it is conveniently insensitive to dust reddening. This quantity can also be used as an estimate of the ionisation state of the gas in the NLR (e.g., \citealt{netzer09,kewley13}).

We now quantify the correlation between the width of the broad H$\alpha$ and the narrow L([OIII])/L(H$\beta$) ratio. To do so one could, in principle, fit each emission line spectrum individually and decompose the narrow and the broad kinematic components. However, as can be seen in \cite{shen11}, the broad emission lines often show complex and asymmetric profiles. As their flux dominates the emission in that spectral region, the characterisation of the (weaker) associated narrow lines can be highly uncertain in such fits, especially when the broad lines become more Lorentzian. This limitation can be alleviated or greatly reduced by averaging a collection of spectra from consecutive objects in the sequence. This leads to better defined and more symmetric line shapes, allowing a more robust estimate of the narrow lines. To estimate the overall trend, we divide the sequence into 35 bins with roughly 55 objects per bin. We note that consistent trends are obtained if we increase the number of bins up to 190 (10 objects per bin) but with an increasing level of scatter. For each group of consecutive objects from the sequence, we compute a median line emission spectrum and perform a broad and narrow line decomposition. To do so, we first fit and subtract the contribution of FeII emission following the approach presented by \citet{kovacevic10}. We model the broad H$\alpha$ and H$\beta$ emission lines as Voigt profiles and the narrow H$\alpha$, H$\beta$, [OIII], and [NII] as Gaussians. We tie all the central wavelengths to the same systemic velocity and we use a single velocity dispersion to describe the narrow lines. The amplitudes of the [NII] and [OIII] doublets are set to their theoretical ratios. In order to obtain a satisfactory representation of the data we find that an additional broader kinematic component is required for the [OIII] emission lines. This component is likely due to AGN-driven winds which are observed in numerous AGN spectra \citep{heckman81,greene05a,komossa08,wylezalek16,baron17b}. To account for this, we add two Gaussians for [OIII]$\lambda$4959\AA\ and $\lambda$5007\AA, with amplitudes set to their theoretical ratio. Finally, as is well known, the broad H$\beta$ emission line shows some asymmetry. Accounting for it has been done in various ways in the literature: some authors include several Gaussian components to fit the broad H$\beta$ lines \citep{shen11,trakhtenbrot12}, others use a broken power-law with an asymmetric slope around the central wavelength which is then convolved with a Gaussian \citep{bisogni17}. In our case, we simply add one Gaussian in the region between the H$\beta$ and the [OIII] lines. We note that while this additional feature allows us to obtain better fits to the data, it does not affect the findings presented in this analysis. We show in Figure \ref{f:stacked_spectra_fitting_example} four examples of the best-fitting profiles to the median spectra. In each row, the panel shows the H$\beta$ and H$\alpha$ regions. The figure shows that our spectral parameterisation provides a description of the data that is sufficiently good for our purposes.
    
Having obtained a best-fit parameterisation for each median spectrum, we estimate the flux of the narrow [OIII] and H$\beta$ emission lines by integrating their best-fit Gaussian profiles. The FWHM of the broad H$\alpha$ is approximated by: $0.5346f_{L} + \sqrt{0.2166f_{L}^{2} + f_{G}^{2}}$, where $f_{L}$ is the FWHM of the Lorentzian and $f_{G}$ is the FWHM of the Gaussian. This analytic form for the FWHM of a Voigt profile has an error which is smaller than the uncertainty involved in the fitting process \citep{olivero77}. 

The dependence of the L([OIII])/L(H$\beta$) line ratio as a function of the FWHM of the broad H$\alpha$ is shown in the right panel of Figure \ref{f:narrow_broad_lines_correlation}. This relation indicates that the narrow L([OIII])/L(H$\beta$) emission line ratio increases as the broad lines become wider over the entire range of H$\alpha$ velocity dispersions. 
First, we focus on AGN-dominated systems with log([OIII]/H$\beta$)$>0.55$. In this range we observe a simple power-law dependence between the two observables. To estimate its best fit parameters we perform a $\chi^{2}$ minimisation, taking into account the uncertainties in both the FWHM of the broad H$\alpha$ and that of the narrow L([OIII])/L(H$\beta$) line ratio. The best-fitting relation is given by:
    \begin{equation}
    \begin{split}
    \label{eq:scaling}
    \log \left( \frac{L(\mathrm{[O_{\rm III}^{\rm narrow}]}) }{ L({\rm H}_\beta^{\rm narrow})} \right)
    =& (0.58\pm0.07)\,\times\,\log \left(\mathrm{\frac{FWHM\,H_\alpha^{\rm broad}}{km/sec} } \right)\\
    &-1.38\,\pm\,0.38\;,
    \end{split}
    \end{equation}
and is marked with a black line in the right panel of Figure \ref{f:narrow_broad_lines_correlation}. This new scaling relation ties the kinematics of the gas clouds in the BLR to the ionisation state of the NLR, such that the NLR gas is more ionised in systems with larger velocity dispersions in their BLR gas. This novel connection shows the existence of a physical process connecting the properties of gas clouds typically a kiloparsec away from the black hole to material gravitationally bound to it on a scale smaller than a parsec. 

The lower-end of the y-axis, i.e. for log([OIII]/H$\beta$)$<0.55$, is dominated by narrow-line Seyfert 1 galaxies (NLS1), which are type~I AGN with \emph{narrow} optical broad lines (FWHM H$\beta \leq 2000$ km/sec) that show large [OIII]/H$\beta$ ratios in their narrow lines (e.g., \citealt{osterbrock85,goodrich89,ardila00,xu07}). One can see a steeper slope of the L([OIII])/L(H$\beta$) - FWHM(broad H$\alpha$) relation in this regime. We then examine whether such a slope is due to star formation in the host galaxy. \citet{sani10} found that such systems show enhanced star formation rates compared to AGN with broader emission lines. Since star formation-related ionisation results in lower L([OIII])/L(H$\beta$) ratios (usually $< 0.5$), this enhanced contribution of star formation to the ionisation of the gas may account for the lower L([OIII])/L(H$\beta$) ratios observed in this regime, resulting in a steeper slope. In appendix \ref{s:sf_contribution}, we analyse a sample of type~II AGN, for which the star formation rate is measured through the continuum emission. We find that the narrow L([OIII])/L(H$\beta$) ratio correlates with the star formation rate for systems with log([OIII]/H$\beta$) $<0.55$, suggesting that it is influenced by star formation in the host galaxy. However, the correlation disappears above log([OIII]/H$\beta$) $\sim 0.55$, suggesting that Eq.~\ref{eq:scaling} is not contaminated by star formation in the host galaxy, and is an intrinsic property of AGN.

\begin{figure}
\includegraphics[width=0.5\textwidth]{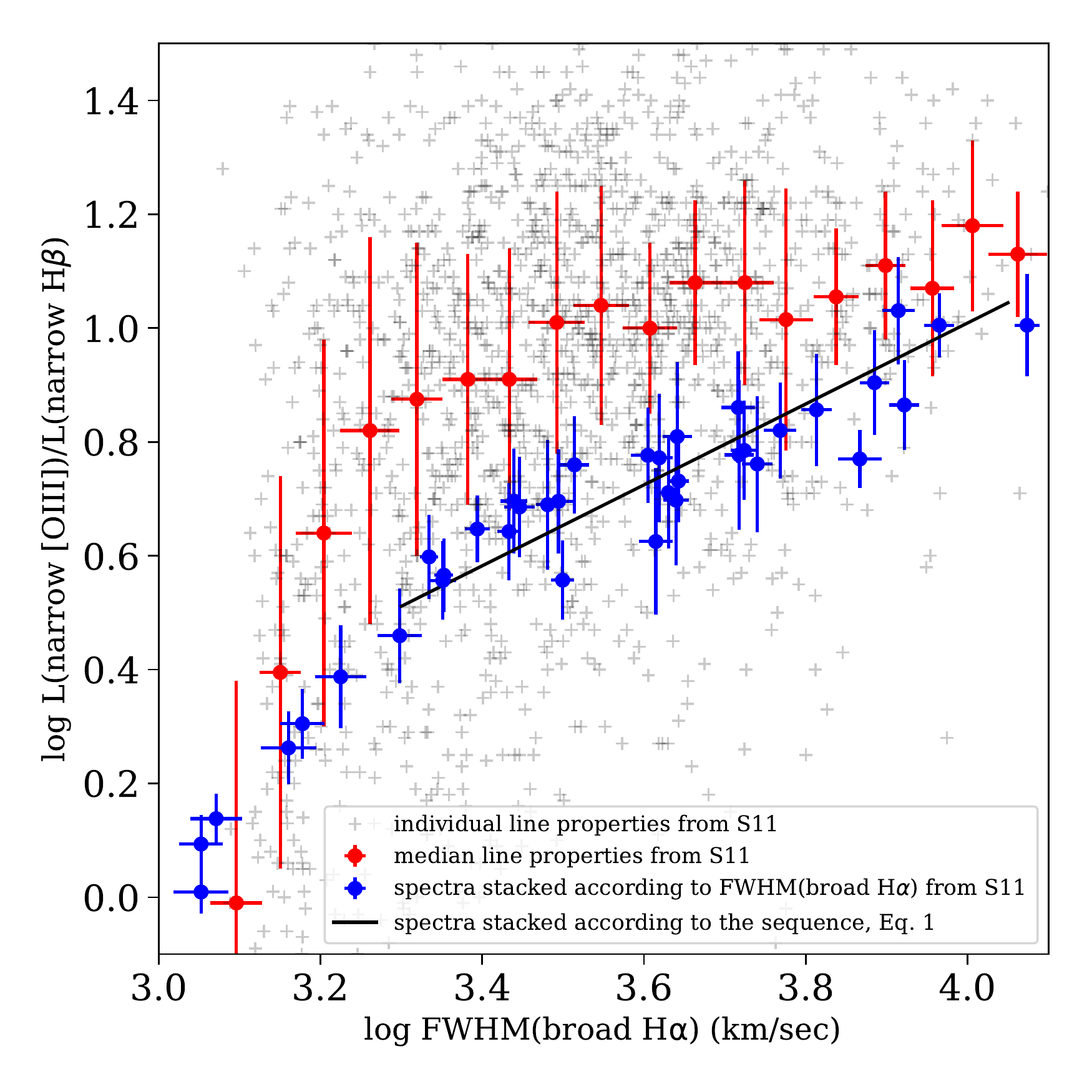}
\caption{The narrow line ratio L([OIII])/L(H$\beta$) as a function of the width of the broad H$\alpha$ for different estimators: individual measurements obtained from the \citet{shen11} catalogue (grey crosses), median values measured in bins of the FWHM of the broad H$\alpha$ (red data points) and values measured from \emph{median spectra} in the same bins (blue crosses). The difficulty in robustly measuring the properties of narrow lines on top of broad lines in individual systems tends to dilute the underlying correlation. For reference, the best-fit relation (Eq.~\ref{eq:scaling}) obtained using the sequence in line shape is shown with black solid line.}
\label{f:comparison_with_shen_et_al}
\end{figure}

Next, we show that the relation shown in Figure \ref{f:narrow_broad_lines_correlation} can also be obtained without the availability of the sequence presented above. To do so we use the measured line parameters provided by \cite{shen11}, more specifically the FWHM of the broad H$\alpha$ and the luminosity of the narrow [OIII] and H$\beta$ emission lines. These values are based on emission line decompositions of individual AGN spectra. In Figure \ref{f:comparison_with_shen_et_al} we show the L([OIII])/L(H$\beta$) narrow emission line ratio versus the FWHM of the broad H$\alpha$ for individual systems (small grey crosses). A lower envelope can be seen. If we combine these measurements using bins of FWHM(broad H$\alpha$) and show the median L([OIII])/L(H$\beta$) line ratio we do detect a correlation between these two parameters, as shown by the large red crosses for which the error bars denote the underlying scatter. These points indicate a significant but relatively shallow dependence for most values of FWHM(broad H$\alpha$). Interestingly, if we use the same bins but we instead estimate median spectra and then measure line properties, i.e. the width of the broad H$\alpha$ line and the line luminosity of [OIII] and the narrow H$\beta$, we find the steeper trend shown with the blue points which is consistent with the best-fit presented in Eq.~1. In other words, we recover our main result but without the use of the initial sequence. The weaker correlation found above illustrates the difficulty in properly estimating narrow line properties on top of broad lines in individual objects. In the presence of noisy, complex and asymmetric broad line H$\beta$ profiles, it appears that the narrow H$\beta$ lines tend to be underestimated.

\subsection{The BPT diagram} 
    
\begin{figure}
\includegraphics[width=0.5\textwidth]{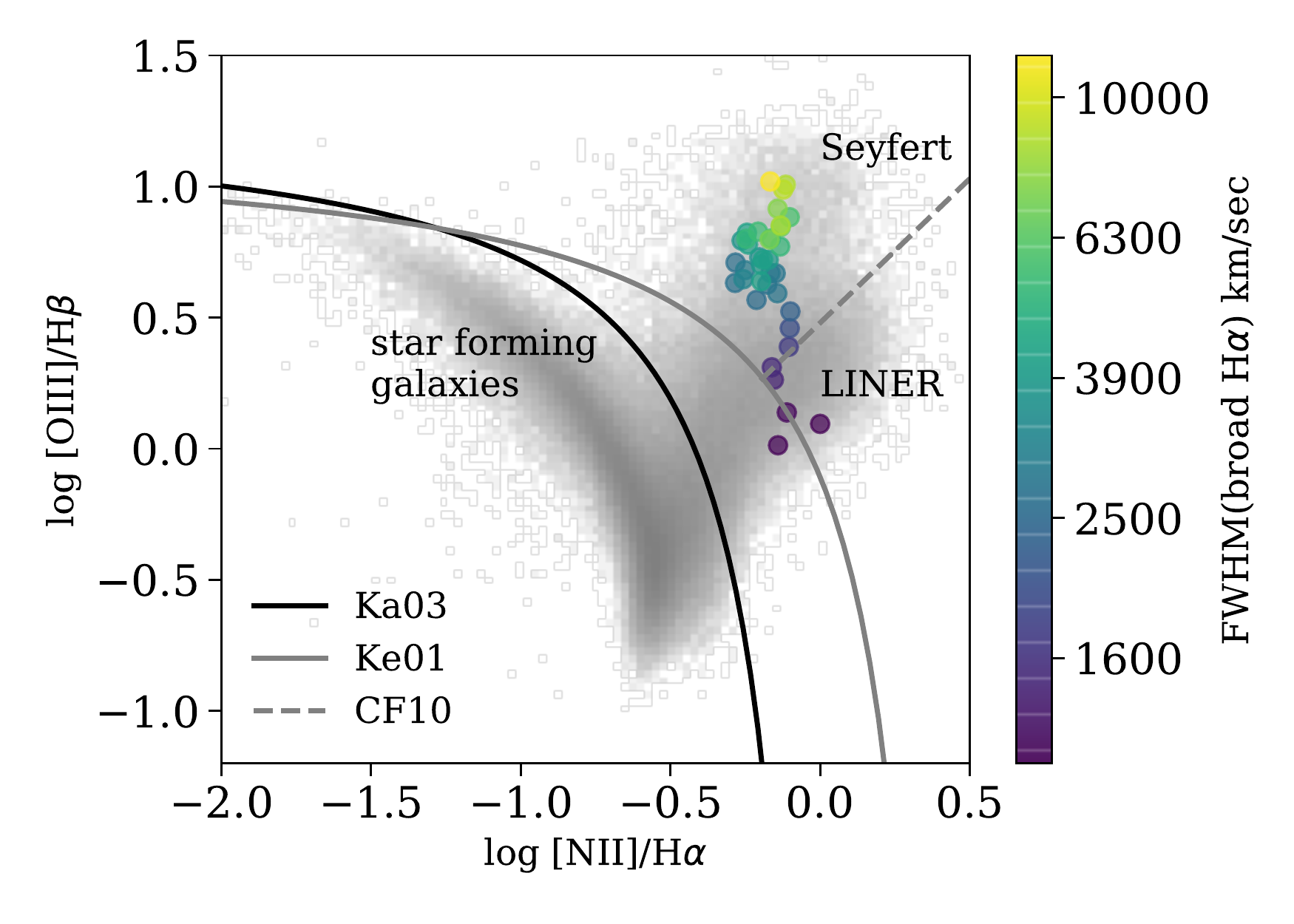}
\caption{The narrow emission line ratios of the stacked AGN spectra on the standard BPT diagram. We show the L([OIII])/L(H$\beta$) versus L([NII])/L(H$\alpha$) emission line ratios, where each measurement is coloured according to the FWHM of the broad H$\alpha$. We mark the extreme starburst line by Ke01 with black, the composite line by Ka03 with grey, and the LINER-Seyfert separation by CF10 with yellow. The background represents the distribution in emission line ratios of the SDSS emission line galaxies.}
\label{f:bpt_diagram_with_fwhm_halpha}
\end{figure}

It is informative to present the correlation between narrow and broad lines in the context of the so called BPT diagram \citep{baldwin81,veilleux87}, shown in Figure~\ref{f:bpt_diagram_with_fwhm_halpha}. This diagram shows the L([OIII])/L(H$\beta$) ratio, which depends primarily on the ionisation parameter, as a function of the L([NII])/L(H$\alpha$) ratio, which depends primarily on metallicity. The combination of these two line ratios provides a diagnostic of the main source of ionising radiation (young stars versus AGN) and gas properties \citep{groves04b,kewley06,kewley13}. The colour-coded points in the figure show the values of L([OIII])/L(H$\beta$), L([NII])/L(H$\alpha$) and the FWHM(broad H$\alpha$) measured from median spectra computed for each of the 35 bins of consecutive objects in the overall sequence presented above. First, we confirm that our objects occupy the Seyfert region in the diagram. Interestingly, we find that the locus given by those two correlations follows the central part of the Seyfert region of the BPT diagram. It informs us that the Seyfert track of the BPT diagram is not only a sequence in the ionisation parameter of the narrow line region, but also a sequence in the kinematics of the gas in the broad line region, indicated by the width of the broad H$\alpha$ line.
    
\section{A black hole mass estimate for type\,2 AGN}\label{s:new_black_hole_mass_estimate}
    
In type~I AGN, black hole mass is estimated using the width of the visible broad emission lines. This is usually referred to as the virial black hole mass or the single epoch mass determination, which relies on the radius-luminosity relation, found from reverberation mapping studies \citep{kaspi00,peterson04,bentz13}. This technique has been applied to various AGN samples \citep{greene06,shen11,trakhtenbrot12}. In contrast, for type~II AGN the accretion disk and the broad line region are obscured by a dusty torus and are thus not visible to us. Since our analysis has revealed a connection between the kinematics of the clouds in the BLR and the ionisation state of the clouds in the NLR, we can now use this correlation to estimate the black hole mass in type~II AGN based only on the narrow lines. According to the unification model \citep{antonucci93, urry95, netzer15}, type~I \& II AGN represent the same physical system, viewed from a different inclination. Thus, the properties of the narrow line region should be similar in the two types of AGN, and the relation we found for type~I should also apply to type~II objects. 
    
We describe the type~II AGN sample that we use in section \ref{s:type2_sample}. We use the relation we find between the narrow and the broad lines, and additional scaling relations from the literature, to estimate the black hole masses of these systems in section \ref{s:bh_mass_type2}. Finally, since in type~II AGN the host galaxy is directly observed, we present scaling relations between the mass of the supermassive black hole and host galaxy properties, such as stellar velocity dispersion and stellar mass, in section \ref{s:scaling_relations}.

\subsection{Type~II AGN sample}\label{s:type2_sample}
    
Using again data from SDSS DR7, we select objects classified as Seyferts based on the classical line-diagnostic diagrams in the [OIII]/H$\beta$ - [NII]/H$\alpha$ plane. To select objects with well-defined line properties, we cross-match the MPA-JHU catalogue \citep{kauff03b, b04, t04} with the OSSY catalogue \citep{oh11} and keep galaxies for which both catalogues report emission lines detected at more than 3$\sigma$. This procedure allows us to minimise cases for which sky line residuals and problematic data affect the proper characterisation of the emission lines \citep{baron17a}. For our analysis, we then use the parameters given in the MPA-JHU catalogue. We select Seyferts as objects which exhibit [OIII]/H$\beta$ emission line ratios which are above the extreme starburst limit (Ke01; \citealt{kewley01}) and use a standard criterion by CF01 \citep{cidfernandes10} to separate Seyferts from LINERs. This leads to a sample of 9\,796 objects. We then estimate the stellar velocity dispersion of each system. To do so, we use pPXF \citep{cappellari12}, a public code for extracting the stellar kinematics and stellar population from absorption line spectra of galaxies. It uses the MILES library \citep{vazdekis10}, which contains single stellar population synthesis models and covers the full range of the optical spectrum with a resolution of FWHM of 2.3\AA. Finally, we make use of stellar mass estimates using the catalogues provided by SI11 \citep{simard11} and M14 \citep{mendel14}. SI11 performed a bulge/disk decomposition for more than a million SDSS galaxies in the $g$ and $r$ bands, improving the SDSS pipeline estimate by considering the two-dimensional point spread function and object deblending. M14 extended their work by performing the bulge and disk decomposition in $u$, $i$, and $z$ bands as well. They derive the total, bulge, and disk stellar masses using broadband spectral energy distribution fitting, with a typical uncertainty of 0.15 dex. We use the total stellar masses measured by M14 in section \ref{s:scaling_relations}.

\subsection{Black hole mass estimation}\label{s:bh_mass_type2}
    
Typically, black hole mass estimation with type~I AGN is done by assuming that the gravity due to the black hole dominates the kinematics of the clouds in the broad line region. One can then estimate the mass dynamically via the scaling $M_{\mathrm{BH}} \propto R \Delta V^{2}$, where the average gas velocity can be obtained using the width of the broad emission lines and the radius is estimated using the radius-luminosity relation defined by reverberation mapping \citep{kaspi00,peterson04,bentz13}. Our goal here is to follow the classical black hole mass estimation method used for type~I AGN (e.g., \citealt{reines15,bentz18}) and apply it to type~II AGN by using the new scaling relation given in Eq.~\ref{eq:scaling} to statistically estimate dynamical information of the (invisible) broad line region.
    
Starting from the virial theorem, one can estimate the mass of a black hole as a function of the radius $R_\mathrm{BLR}$ of the broad line region and the gas velocity dispersion, typically through the broad H$\beta$ line:
    \begin{equation}
    \label{eq:virial}
    	{M_{\mathrm{BH}} = \epsilon \; \Big( \frac{R_{\mathrm{BLR}}\,\mathrm{FWHM(H\beta)^{2}}}{G} \Big)}\;,
    \end{equation}
    modulated by a scaling factor $\epsilon$ which depends on the geometry of the broad line region and its apparent inclination, both of which are usually unknown. A radius-luminosity relation constrained by reverberation mapping is then used to infer $R_\mathrm{BLR}$. We follow the procedure done for type~I AGN \citep{reines15} and use the modified radius-luminosity relation \citep{bentz13}:
    \begin{equation}
    \label{eq:rl_relation}
    	{\log \Big( \frac{R_{\mathrm{BLR}}}{\mathrm{light-days}} \Big) = 1.555 + 0.542 \log \Big( \frac{L_{\mathrm{5100}}}{10^{44}\, \mathrm{erg/s}} \Big)},
    \end{equation}
    where $L_{\mathrm{5100}}$ is the AGN continuum luminosity at 5100\AA. Finally, to express this relation in terms of H$\alpha$ width, we use the empirical correlation between the widths of the broad Balmer lines \citep{greene05b}:
    \begin{equation}
    \label{eq:hahb}
    \mathrm{FWHM(H\beta) = 1.07\times 10^{3} \Big( \frac{FWHM(H\alpha)}{10^3\, km/s} \Big)^{1.03}\, km/s}\;.
    \end{equation}
    Combining the above equations, we obtain one of the standard expressions used to estimate black hole masses in type~I AGN:
    \begin{eqnarray}
    \label{eq:bhm}
     \log \Big( \frac{M_{\mathrm{BH}}}{M_{\odot}} \Big) &=& \log \epsilon + 6.90 + 0.54 \log \Big( \frac{L_{5100}}{10^{44} \, \mathrm{erg/s}} \Big)  \nonumber\\
     && + 2.06 \log \Big( \frac{\mathrm{FWHM(H\alpha)}}{10^3\, \mathrm{km/s}} \Big)
    \end{eqnarray}
    As noted by several studies (e.g., \citealt{peterson04,collin06,denney10,greene10,barth11}), this black hole mass estimation is indirect, and is subject to various uncertainties.

To estimate the black hole mass in type~II Seyferts, we use the classical scaling given in Eq.~\ref{eq:bhm} as our starting point. First, we need an estimate of the scaling factor $\epsilon$. Typically, a generic value is assumed by calibrating reverberation-based black hole masses to the local $M_{\mathrm{BH}}-\sigma_{*}$ relation \citep{grier13,woo15}. In order to best compare our results to previous estimates using type~I AGN, we follow \citet{reines15} and use $\epsilon=1.075$, corresponding to a mean virial factor of $<f> = 4.3$. To estimate the AGN continuum luminosity at 5100\AA, we can use its measured correlation with the bolometric one. Using a standard bolometric correction \citep{runnoe12} and after accounting for anisotropy in the accretion disk emission \citep{nemmen10}, one obtains:
    \begin{equation}
    \log L_{5100} = 1.09 \times \log L_{\mathrm{bol}} - 5.23\;.
    \end{equation}
    A bolometric luminosity estimate can then be inferred statistically from the narrow H$\beta$ and [OIII] luminosities. It is found that \citep{netzer09}:
    \begin{equation}
    \label{eq:bol}
    	{\log L_{\mathrm{bol}} = \log L(\mathrm{H}\beta) + 3.48 + \mathrm{max \Big[ 0, 0.31\big( \log \frac{[OIII]}{H\beta} -0.6 \big) \Big] }}\;.
    \end{equation}
Finally, we correct the luminosity estimate for dust extinction. To do so we use the narrow H$\alpha$ and H$\beta$ emission lines to estimate the dust reddening in the NLR. Assuming a dusty screen, case-B recombination, a gas temperature of $10^4$ K \citep{osterbrock06} and the CCM extinction law \citep{cardelli89}, the dust reddening correction is given by:
    \begin{equation}
    \label{eq:dust}
    \mathrm{E}(B-V) = 2.33 \log \Big[ \frac{\mathrm{(H\alpha/H\beta)_{obs}}}{2.85} \Big]\;.
    \end{equation}
    We use it to estimate the extinction-corrected H$\beta$ luminosity (and therefore $L_{5100}$ according to Eq.~\ref{eq:bol}) for each system. Finally, we can estimate the last term of Eq.~\ref{eq:bhm}, the FWHM of the broad H$\alpha$ line, by inverting Eq.~\ref{eq:scaling} and using the measured values of the narrow line ratio L([OIII])/L(H$\beta$).

    \underline{To summarise:} using the new scaling relation given in Eq.~\ref{eq:scaling}, an estimate of the luminosity $L_{5100}$ obtained through its correlation with the bolometric luminosity through the H$\beta$ and O[III] lines, and the standard line-based dust extinction correction, we can estimate all the terms required for black hole mass estimation, as given by Eq.~\ref{eq:bhm}. We can therefore estimate black hole masses for type~II Seyferts using only the luminosities of the narrow H$\alpha$, H$\beta$ and [OIII] lines. As we will show below, this indirect estimate is subject to an uncertainty of at least 0.5 dex but it can be used with thousands of objects from existing datasets. We note that, if X-ray measurements are available, they can then be used to estimate $L_{5100}$ more directly and with a smaller scatter \citep{netzer06,maiolino07, lusso12,bongiorno14,koss17}. In this case, only the narrow line ratio L([OIII])/L(H$\beta$) is needed to obtain a black hole mass estimate. Given a fixed observed wavelength range, this will allow us to use this new black hole mass estimate up to even higher redshift as the visibility of the H$\alpha$ line is no longer needed.

\subsection{Relations to galaxy properties}
\label{s:scaling_relations}
    
Beyond the local universe, black hole masses are estimated only in type~I AGN. However, the host galaxy properties are more difficult to measure, since the radiation originating from the accretion disk outshines the stars in the galaxy and dominates the optical wavelength range. Different techniques have been used to separate the AGN from its host galaxy \citep{greene06_vel,peng06,merloni10,bongiorno14,shen15}, leading to hundreds of type~I AGN systems for which both the black hole mass and the host properties are measured. They are generally limited to $z \le 0.1$. These studies reveal a correlation between black hole mass and host properties, but the corresponding best-fit relations show a range of values, sometimes different from those obtained with dynamic black hole mass estimates in passive galaxies \citep{greene06,reines15}. It is still unclear whether these correlations apply only to classical bulges or whether different galaxy morphologies obey different black hole-host relations \citep{greene08,greene10,xiao11,bennert15,reines15}. 
    
\begin{figure*}
\includegraphics[width=0.49\textwidth]{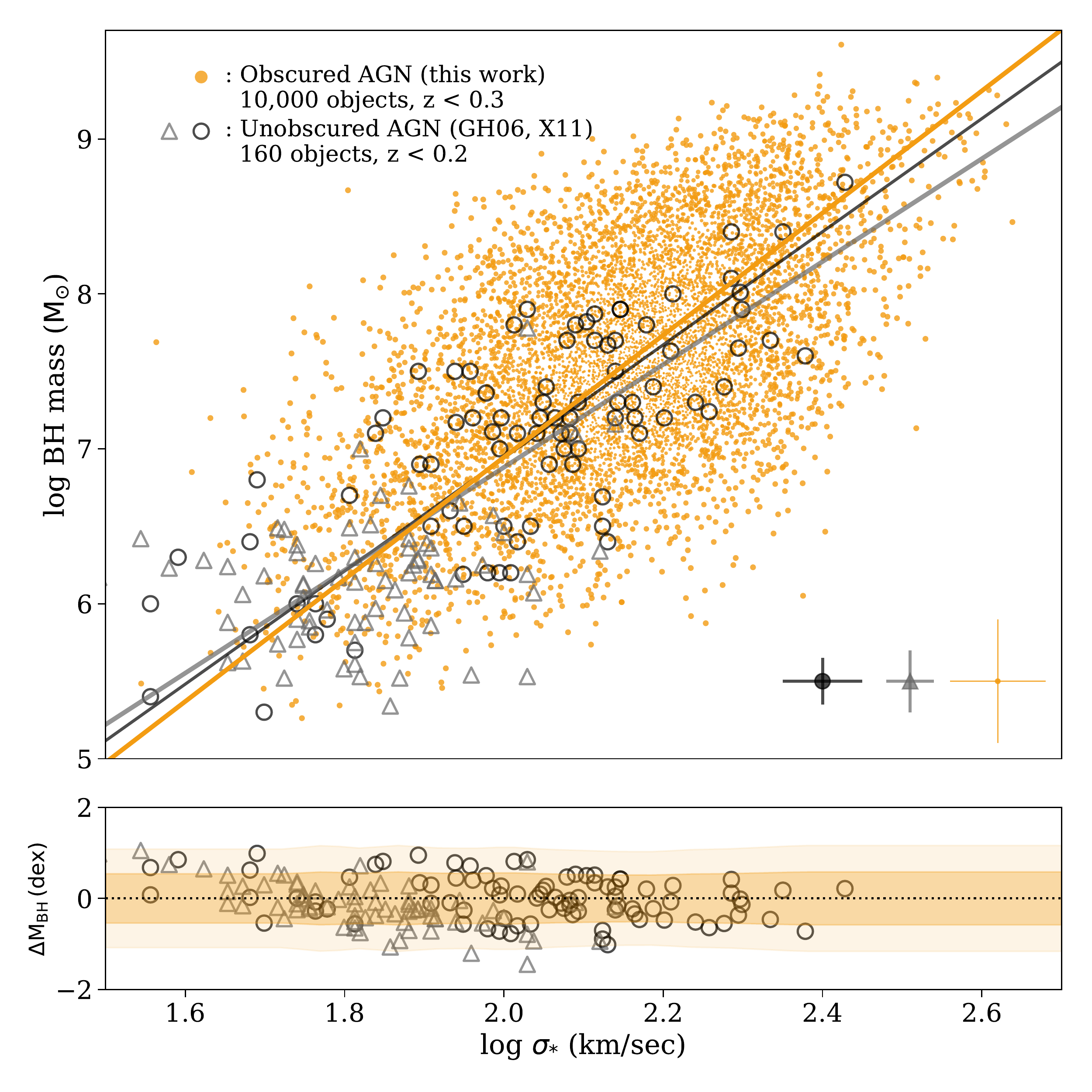}	
\includegraphics[width=0.49\textwidth]{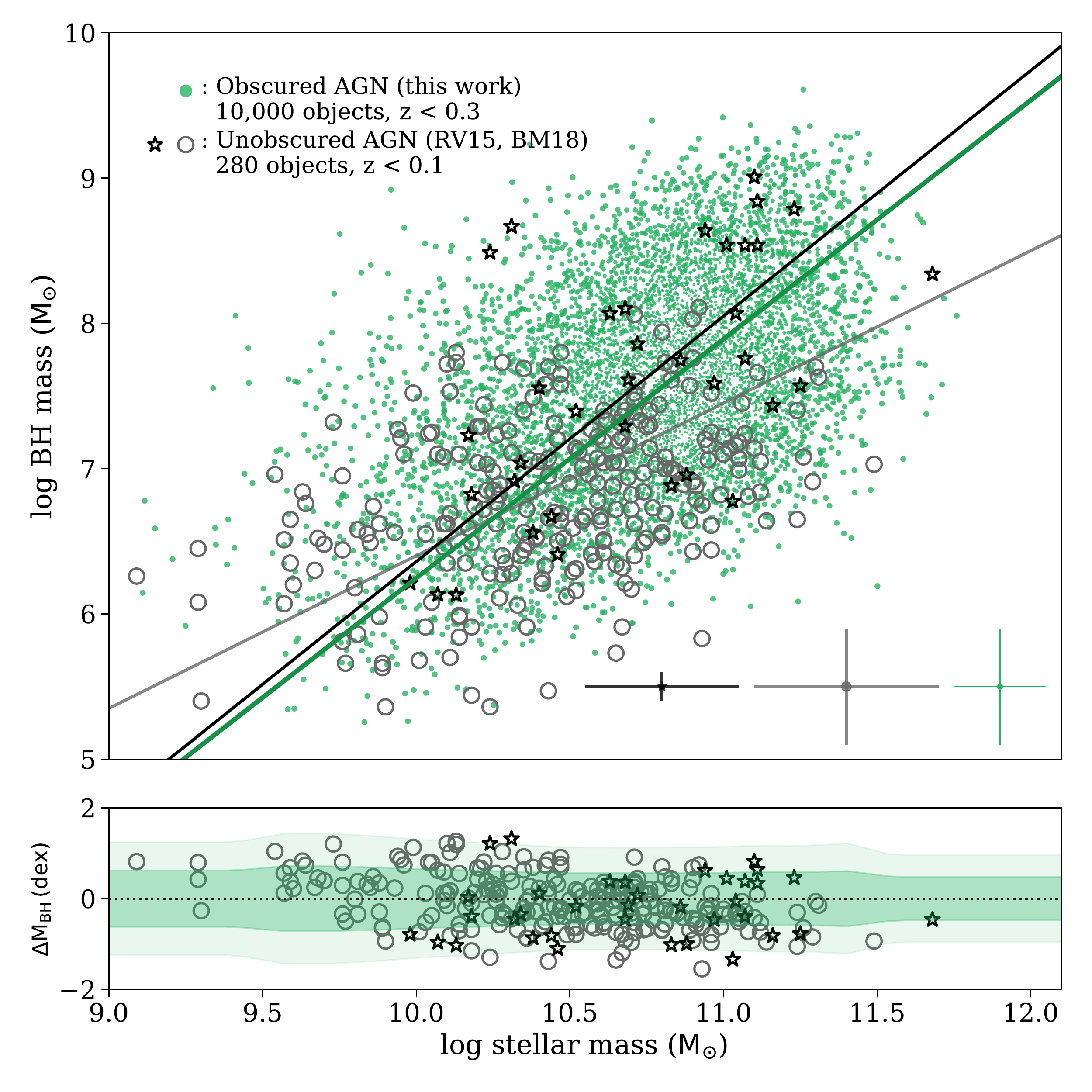}
\caption{\textbf{The relation between black hole mass and host galaxy properties in the two types of AGN.} Left panel: black hole mass estimates for about $10\,000$ \emph{obscured} type~II AGN based on the narrow line ratio introduced in this work, as a function of galaxy stellar velocity dispersion, and best-fit relation (orange). In black and grey we show estimates for unobscured type~I AGN from GH06 \& X11 for which black hole mass estimates are derived from the width of the broad lines. Bottom left panel: residuals around the best-fitting relations in obscured and unobscured sources. The orange bands denote the $1$ and $2\sigma$ contours for obscured AGN. This shows that the two different black hole mass estimates are subject to comparable scatter. Right panels: similar to the left panels, with black hole mass as a function of total stellar mass (green). In black and grey we show estimates for unobscured type~I AGN from RV15 \& BM18.}
\label{f:m_sigma_m_mbulge_final_version}
\end{figure*}
	
We now use the new black hole mass estimation described in section \ref{s:bh_mass_type2} to examine the $\mathrm{M_{BH}-\sigma_{*}}$ and $\mathrm{M_{BH}-M_{*}}$ relations for the sample of about $10\,000$ type~II AGN previously introduced. We have applied our black hole mass estimate for objects as a function of stellar velocity dispersion and stellar mass and we show the results in Figure \ref{f:m_sigma_m_mbulge_final_version}. They show that the black hole mass of type~II AGN is strongly correlated to stellar velocity dispersion and stellar mass. To quantify these trends we perform maximum likelihood estimation. To do so, we need to take into account the strongly non-uniform mass distribution of the SDSS Seyfert sample: as shown in Figure \ref{f:m_sigma_m_mbulge_final_version}, a large fraction of the data points cluster around $\log \sigma_*/({\rm km/s})\sim2.2$ and $\log M_{*}/M_\odot \sim10.8$ due to the selection properties of the sample. We label $y=\mathrm{\log M_{BH} / M_{\odot}}$ and $x=\mathrm{\log \sigma_{*}/(km/sec)}$ or $\mathrm{\log M_{\rm *} / M_{\odot}}$, and model the expected trends as $y = ax + b + \epsilon$ where the $x$ values follow approximately a Gaussian distribution with parameters $x \sim \mathcal{N}(\mu_{x}, \sigma_{x}^2)$ estimated from the data and $\epsilon \sim \mathcal{N}(0, \sigma_{y}^2)$ describes the scatter in $y$ and for which we set $\sigma_y=0.3\,{\rm dex}$. The log-likelihood is given by:
    \begin{equation}
    \begin{split}
    & \log L(a, b, \sigma_{y}^{2}, \mu_{x}, \sigma_{x}^{2})  = \log \prod_{i=1}^{N} p(y_{i} | a, b, \sigma_{y}^{2}, \mu_{x}, \sigma_{x}^{2})\\
    & = \sum_{i=1}^{N} \log \int_{-\infty}^{\infty} p(y_{i} | a, b, \sigma_{y}^{2}, x) f(x | \mu_{x}, \sigma_{x}^{2}) dx\nonumber \\ 
    & = \sum_{i=1}^{N} \log \int_{-\infty}^{\infty} \frac{{\rm d}x}{2 \pi \sigma_x \sigma_y} \exp \Big[ -\frac{(x - \mu_{x})^2}{2 \sigma_{x}^2} - \frac{[y_i - (ax + b)]^2}{2 \sigma_y^2} \Big]\;.
    \end{split}
    \end{equation}
    We maximise it to estimate the parameters $a$ and $b$ and to obtain the best fit estimates in the $\mathrm{M_{BH}-\sigma_{*}}$ and $\mathrm{M_{BH}-M_{*}}$ relations for type~II AGN:
    \begin{equation}
    \mathrm{ \log \frac{M_{BH}}{M_{\odot}}} =
    \left\{
        \begin{array}{ll}
    \vspace{0.2cm}
            \mathrm{ (3.94 \pm 0.12) log \left(\frac{\sigma_*}{200\, km/s}\right) + (8.13 \pm 0.07) }\\
            \mathrm{ (1.64 \pm 0.18) log \left(\frac{M_{*}}{10^{11}\, M_{\odot}}\right) + (7.88 \pm 0.13) }
        \end{array}
    \right.
    \label{eq:m-sigma-m-bulge}
    \end{equation}
    These best fit relations are shown in Figure \ref{f:m_sigma_m_mbulge_final_version}. In order to meaningfully compare our results to those obtained in type~I AGN, we choose studies for which the host properties were obtained using methods as close as possible to ours. Namely, we compare our results to those presented by GH06 \citep{greene06}, X11 \citep{xiao11}, RV15 \citep{reines15}, and BM18 \citep{bentz18}\footnote{
    GH06 present the $\mathrm{M_{BH}-\sigma_{*}}$ relation for a sample of 88 type~I AGN. X11 extend the $\mathrm{M_{BH}-\sigma_{*}}$ relation to lower black hole masses \citep{greene07}. Both of these studies estimate the black hole mass with a virial factor that corresponds to spherical BLR, which is different from what we use. Furthermore, these studies are based on an older radius-luminosity relation. Thus, we re-compute their black hole masses according to our assumed virial factor and the updated radius-luminosity relation.  RV15 study the relation between the black hole mass and the host galaxy mass, and since our black hole mass estimation follows theirs, we do not re-measure their masses. We note that since RV15 study the correlation of the black hole mass with the \emph{total} stellar mass in the galaxy, we present the relation with the total stellar mass and not the bulge mass. BM18 present a black hole mass - bulge mass and BH mass - total stellar mass relations for 37 AGN with reverberation-based black hole masses, adopting a similar virial factor to the one we use, and using two different mass-to-light ratios. We show the BH mass - total stellar mass relation by BM18, with the mass-to-light ratio by \citet{bell01}. Although more recent, we chose not to compare our results to those presented by BE15 \citep{bennert15} since those use spatially-resolved spectroscopy and measure the velocity dispersion at the effective radius, while we measure the stellar velocity dispersion within the SDSS 3'' fibre.
    }. Interestingly, we find a remarkable agreement in the trends seen for the black hole masses of both types of AGN, as a function of stellar velocity dispersion and total stellar mass. In particular, our $\mathrm{M_{BH}-\sigma_{*}}$ relation is also consistent with more recent results that are based on reverberation-based black hole masses by \citet{woo15} and on IFU-based velocity dispersions by \citet{bennert15}. These consistencies also provide additional support to the new black hole mass estimation method introduced in section \ref{s:bh_mass_type2}. We also point out that while the best-fitting $\mathrm{M_{BH}-M_{*}}$ relation by RV15 is significantly different from what we deduce, their measurements show consistent distribution to ours for $\mathrm{M_{BH}} < 10^{8}\,\mathrm{M_{\odot}}$. We thus suggest that the shallow slope found by RV15 is related to the small dynamical range of black hole masses in their sample.
    
\begin{figure*}
\includegraphics[width=0.49\textwidth]{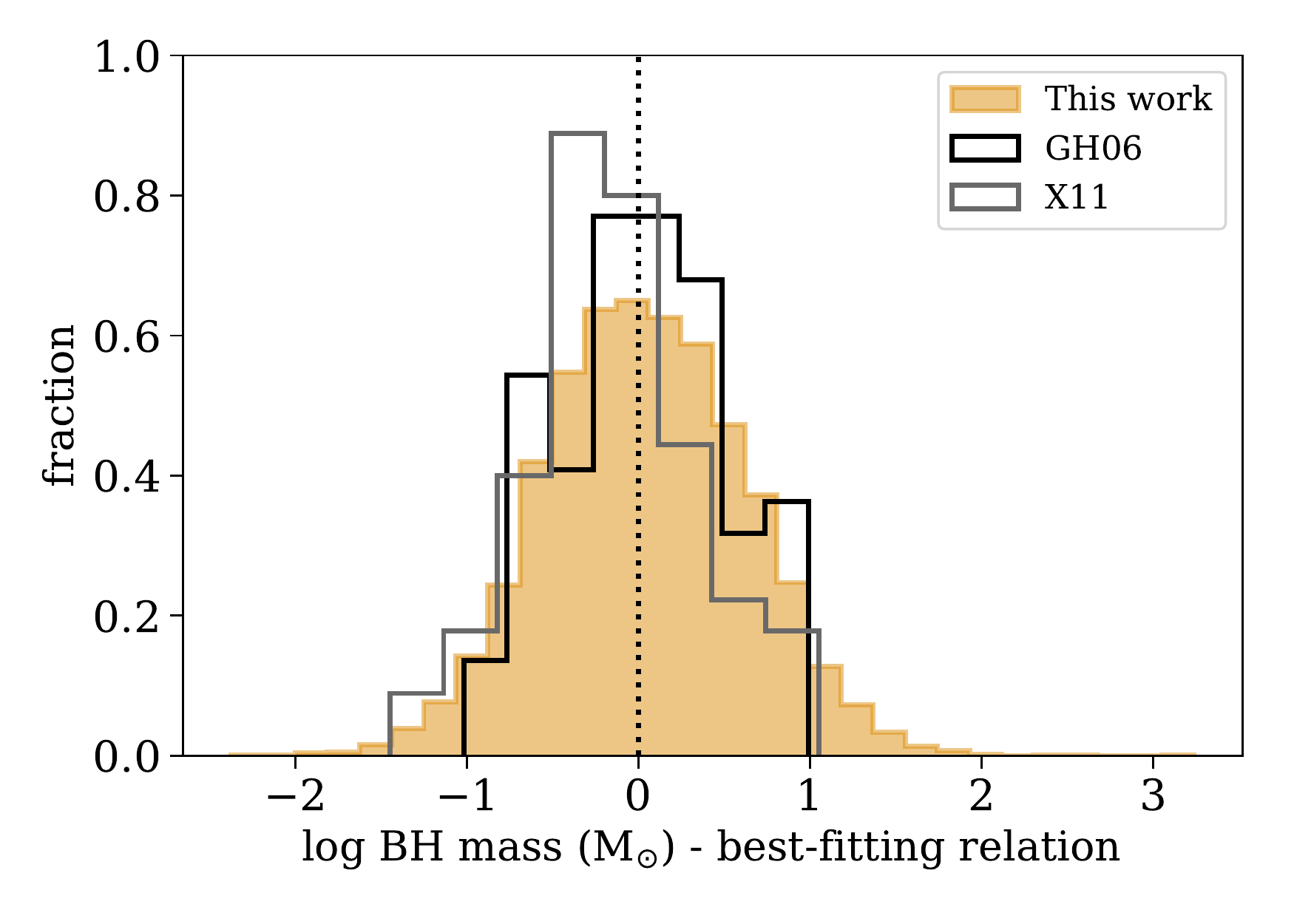}	
\includegraphics[width=0.49\textwidth]{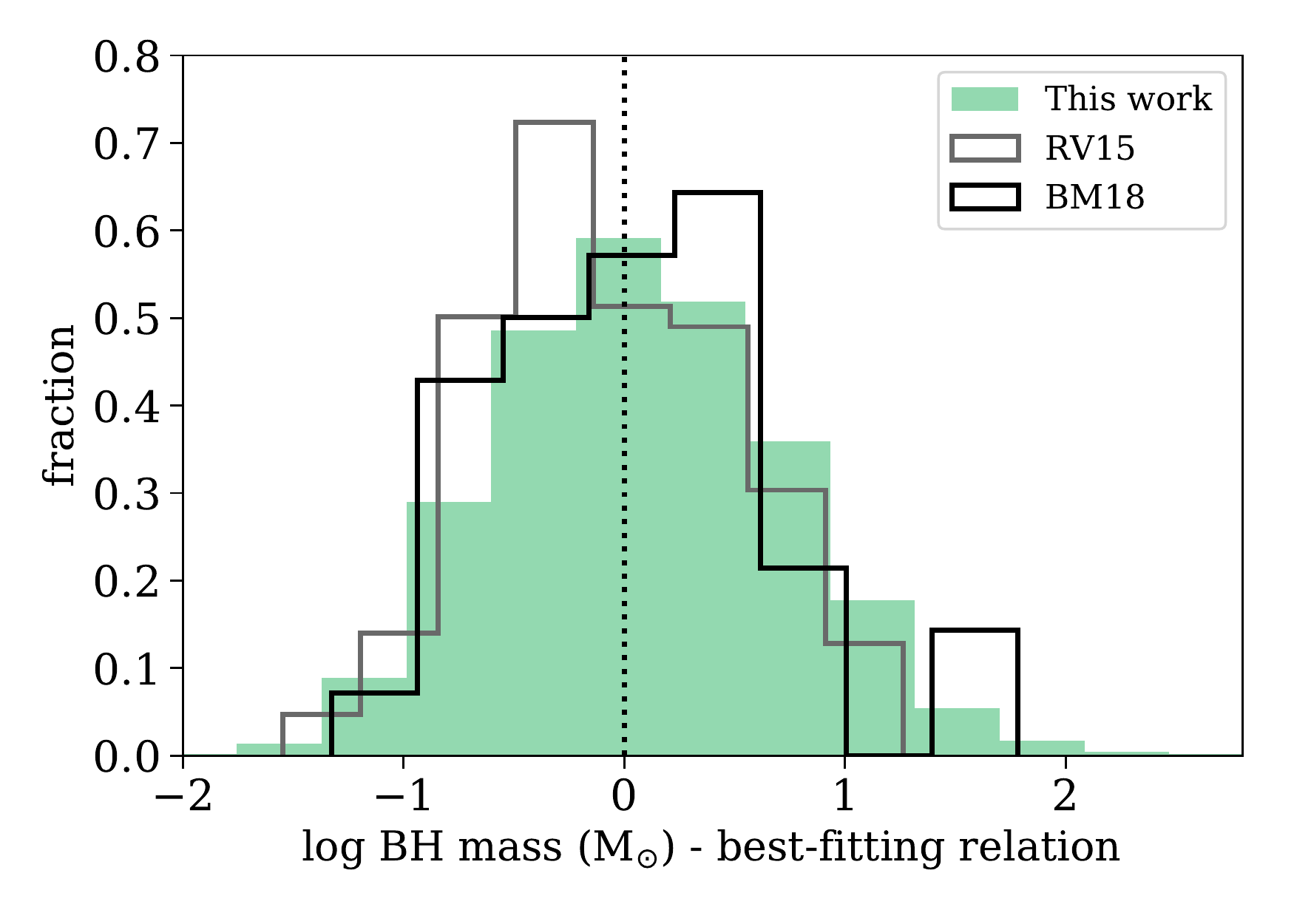}
\caption{Distribution of residuals of the black hole - host galaxy relations in the two types of AGN. The left panel shows the residuals of the $\mathrm{M_{BH}-\sigma_{*}}$ relation in type~II AGN (orange), compared to the relation in type~I AGN by GH06 and X11 (grey and black). The right panel shows the residuals of the $\mathrm{M_{BH}-M_{*}}$ relation in type~II AGN (green), compared to that found in type~I AGN by RV15 and BM18 (grey and black).}
\label{f:m_sigma_m_mbulge_scatter_around_relations}
\end{figure*}	

The errors in the best-fit parameters quoted in Eq.~\ref{eq:m-sigma-m-bulge} were obtained by bootstrapping the selected AGN sample. While they show that the best-fit slope of each relation is well constrained for the specific sample we used, a more informative and more generic way to characterise the precision of the black hole mass estimates is to measure the level of scatter around the best-fitting relations. This is shown in the lower panels of the two figures, and in Figure \ref{f:m_sigma_m_mbulge_scatter_around_relations}. Interestingly, although we used additional scaling relations to estimate the black hole mass, the scatter in the relations for type~II AGN is only slightly larger than the scatter for the type~I AGN. We measure a scatter of 0.33/0.44 dex for the $\mathrm{M_{BH}-\sigma_{*}}$/$\mathrm{M_{BH}-M_{*}}$ relations using the median absolute deviation of the residuals, and 0.45/0.55 dex using the standard deviation. We note that varying the selection of bulge-to-disk ratio does not appear to affect the scatter of these relations. We use the bulge and disk decomposition by SI11 to select systems which are classical bulges and find no reduction in the scatter of the $\mathrm{M_{BH}-\sigma_{*}}$ and $\mathrm{M_{BH}-M_{bulge}}$ relations. Furthermore, we point out that selecting systems with a bulge-to-total stellar mass ratio larger than 80\% does not reduce the observed scatter.
To summarise, the black hole mass estimate introduced in this work is shown to perform as reliably and with a comparable level of scatter as the classical estimate used in non-obscured AGN and based on broad emission lines. 

\section{Discussion}\label{s:disc}

\subsection{Evaluation of the method}

In section \ref{s:new_black_hole_mass_estimate} we proposed a method to estimate black hole masses in type~II AGN. This method relies on the correlation we discovered between narrow and broad line properties in a sample of type~I AGN. The type~I AGN sample used in this study consists of black holes with masses in the range $10^{7.5}\,\mathrm{M_{\odot}} < \mathrm{M_{BH}} < 10^{9.5}\,\mathrm{M_{\odot}}$, while the black hole masses estimated in the type~II AGN sample extend to $\mathrm{M_{BH} \sim 10^{6}\,\mathrm{M_{\odot}}}$ (see e.g., Figure \ref{f:m_sigma_m_mbulge_final_version}). This is due to the systematically lower AGN bolometric luminosity in the type~II AGN sample and is related to the different selection criteria of the two samples. When estimating black hole masses, we point out that our type II AGN selection ensures that we are only considering the range $0.55 < $ log([OIII]/H$\beta$) $< 1.05$, for which Eq.~\ref{eq:scaling} was estimated. The lower black hole masses estimates on the type~II AGN sample relies on the assumption that the scaling relations presented in section~\ref{s:scaling_relations} and Eq.~\ref{eq:scaling} are valid to $\mathrm{M_{BH} \sim 10^{6}\,\mathrm{M_{\odot}}}$. Given that Figure \ref{f:m_sigma_m_mbulge_final_version} shows that our black hole mass estimates are consistent with those found in type~I AGN even in the low mass regime, it suggests that this assumption is likely to be valid. 

We now discuss possible tests that could verify the method presented in this study. First, it should be pointed out that individual type~I AGN cannot be used for that purpose, i.e., by comparing the virial black hole mass obtained from the broad Balmer lines to the black hole mass that is based on narrow L([OIII])/L(H$\beta$) ratio. As explained above, the complexity of emission line shapes encountered in individual spectra prevents reliable estimates of the narrow Balmer lines. As shown in Figure~\ref{f:comparison_with_shen_et_al} and related text, the L([OIII])/L(H$\beta$) ratio obtained through emission line decomposition of individual sources is highly uncertain and often largely overestimated. We also argue that ``changing-look" AGN, which are systems that transition from type~I to type~II (or vice versa) during a period of several years (e.g., \citealt{lamassa15, stern18}), cannot be used to verify our method. Studies suggest that the changing-look phenomenon cannot be attributed to a simple change in the distribution of obscuring material around the black hole, and might be related to a different physical mechanism (e.g., \citealt{lamassa15, stern18}). Therefore, these systems might not represent the typical AGN. 

To test our indirect black hole mass estimate, we propose to use obscured type~II AGN that display broad Paschen lines in near infrared wavelengths (e.g., \citealt{kim10, lafranca15}). Since such systems do not show broad Balmer emission lines, the narrow L([OIII])/L(H$\beta$) can be estimated robustly in individual spectra. A virial black hole mass, estimated through the broad Paschen lines, could then be compared to the narrow line-based black hole mass. Such an approach would be valuable in quantifying the scatter of the black hole mass estimate and possible systematics of the method presented here.

\subsection{The evolution of black hole - host galaxy scaling relation throughout cosmic time}

The $\mathrm{M_{BH}-\sigma_{*}}$ relation in the local universe was likely established during the AGN phase of a galaxy's lifetime. Since the majority of black hole mass was assembled at high redshifts \citep{yu02}, the strongest evidence for $\mathrm{M_{BH}-\sigma_{*}}$ relation evolution will come from the cosmic epoch of $z \sim 1 - 2$. There have been extensive observational efforts in the past decade to probe this possible redshift evolution, with Seyfert-1 galaxies in which the AGN is less dominant \citep{treu04, woo06, treu07, woo08, bennert10, bennert11}, gravitationally-lensed quasars \citep{peng06, peng06b}, using narrow emission lines as surrogates of the velocity dispersion \citep{salviander07}, and detailed AGN-host decompositions \citep{merloni10, shen15}. Only recently, \citet{shen15} presented a large sample with a large enough dynamical range to clearly detect a correlation between single epoch black hole mass and stellar velocity dispersion to $z \sim 1$. In addition to the uncertainties and systematics present in low redshift samples, in high redshift samples, the host bulge and AGN are typically unresolved, and the single epoch mass determination suffers from additional uncertainties (see \citealt{shen13, shen15}). Furthermore, state-of-the-art methods are currently limited to systems with a non-negligible host galaxy contribution (e.g., host fraction of 0.05 at $\lambda \sim 4200$\AA; \citealt{shen15}), which might bias the observed high-redshift relations.

The method presented in this study can be combined with state-of-the-art measurements in type~I AGN to reduce possible biases and systematics in high redshift scaling relations. Since our method provides an estimate of the black hole mass in type~II AGN, it can be used to probe systems in which the AGN are significantly more luminous than their hosts, and thus probe more luminous AGN with more massive black holes. Furthermore, since our method does not require a significant AGN contribution, it might probe less luminous AGN with less massive black holes than those currently probed. Both of these ranges in parameter space were hardly explored so far. Obviously, the relation between the width of the broad H$\alpha$ line and the narrow line ratio L([OIII])/L(H$\beta$) might evolve with redshift, AGN bolometric luminosity, and Eddington ratio. It is important to probe this relation as a function of redshift and for different AGN luminosities and accretion rates to understand the full range in which it can be used.

\section{Summary}\label{s:concs}

In this work, we looked for global trends in the optical spectra of type~I AGN, for which line shapes are often challenging to quantify. We used the \textit{Sequencer}, a novel algorithm we have designed to find sequences in complex datasets. The method re-orders objects in a dataset according to their similarity measured in a multi-scale manner. We applied this technique to a sample of $\sim$2\,000 type~I AGN spectra, and found a one-dimensional trend in the width and the shape of the broad Balmer lines, together with variations in the strength of the narrow lines. Despite the apparent complexity of individual objects, overall, the emission lines of AGN spectra appear to follow a well-defined sequence. In particular, we discovered a correlation between the narrow L([OIII])/L(H$\beta$) line ratio, which traces the ionisation state of the clouds in the narrow line region, with the FWHM of the broad H$\alpha$, which traces the average kinematics of clouds in the broad line region. This novel connection shows the existence of a physical process connecting the properties of gas clouds typically a kiloparsec away from the black hole to material gravitationally bound to it on a scale smaller than a parsec.

From an empirical point of view, this link between broad and narrow lines brings the missing piece required to estimate black hole masses in obscured type~II AGN. We presented a method to estimate the black hole mass in type~II AGN, which relies on the single epoch mass determination method, on the newly discovered L([OIII])/L(H$\beta$) - FWHM(broad H$\alpha$) correlation, and on additional scaling relations from the literature. The method requires only the narrow [OIII], H$\beta$, and H$\alpha$ emission lines, and we applied it on a sample of $\sim$10\,000 low redshift type~II AGN. Since in type~II AGN the accretion disk is obscured by the dusty torus, the host stellar light is robustly detected. We measured the stellar velocity dispersion and used publicly-available stellar mass measurements to present, for the first time, black hole - host galaxy scaling relations in this sample of 10\,000 obscured type~II AGN. We found a good agreement between the $\mathrm{M_{BH}-\sigma_{*}}$ and $\mathrm{M_{BH}-M_{*}}$ relations found in type~I \& II AGN, with a comparable level of scatter, suggesting that the method presented in this work performs as reliably as the classical estimate used in non-obscured type~I AGN.

\section*{Acknowledgments}
We are grateful to H. Netzer for valuable discussions regarding this work.
We thank T. Heckman, J. Krolik, N. Lubelchick, D. Poznanski, J. X. Prochaska, 
I. Reis, J. Silk, and B. Trakhtenbrot for comments and suggestions that helped improve this manuscript. 
This project is supported by the generosity of Eric and Wendy Schmidt by recommendation of the Schmidt Futures program.
The spectroscopic analysis was made using IPython \citep{perez07}. This research made use of Astropy,\footnote{http://www.astropy.org} a community-developed core Python package for Astronomy \citep{astropy13, astropy18}.

This work made use of SDSS-III\footnote{www.sdss3.org} data. Funding for SDSS-III has been provided by the Alfred P. Sloan Foundation, the Participating Institutions, the National Science Foundation, and the U.S. Department of Energy Office of Science. SDSS-III is managed by the Astrophysical Research Consortium for the Participating Institutions of the SDSS-III Collaboration including the University of Arizona, the Brazilian Participation Group, Brookhaven National Laboratory, Carnegie Mellon University, University of Florida, the French Participation Group, the German Participation Group, Harvard University, the Instituto de Astrofisica de Canarias, the Michigan State/Notre Dame/JINA Participation Group, Johns Hopkins University, Lawrence Berkeley National Laboratory, Max Planck Institute for Astrophysics, Max Planck Institute for Extraterrestrial Physics, New Mexico State University, New York University, Ohio State University, Pennsylvania State University, University of Portsmouth, Princeton University, the Spanish Participation Group, University of Tokyo, University of Utah, Vanderbilt University, University of Virginia, University of Washington, and Yale University. 

\bibliographystyle{mn2e}
\bibliography{ref}

\appendix

\section{Details of the method}

\subsection{Relation between the sequence and derived AGN properties}
\label{s:correlations_with_sequence_index}

\begin{figure*}
\includegraphics[width=0.9\textwidth]{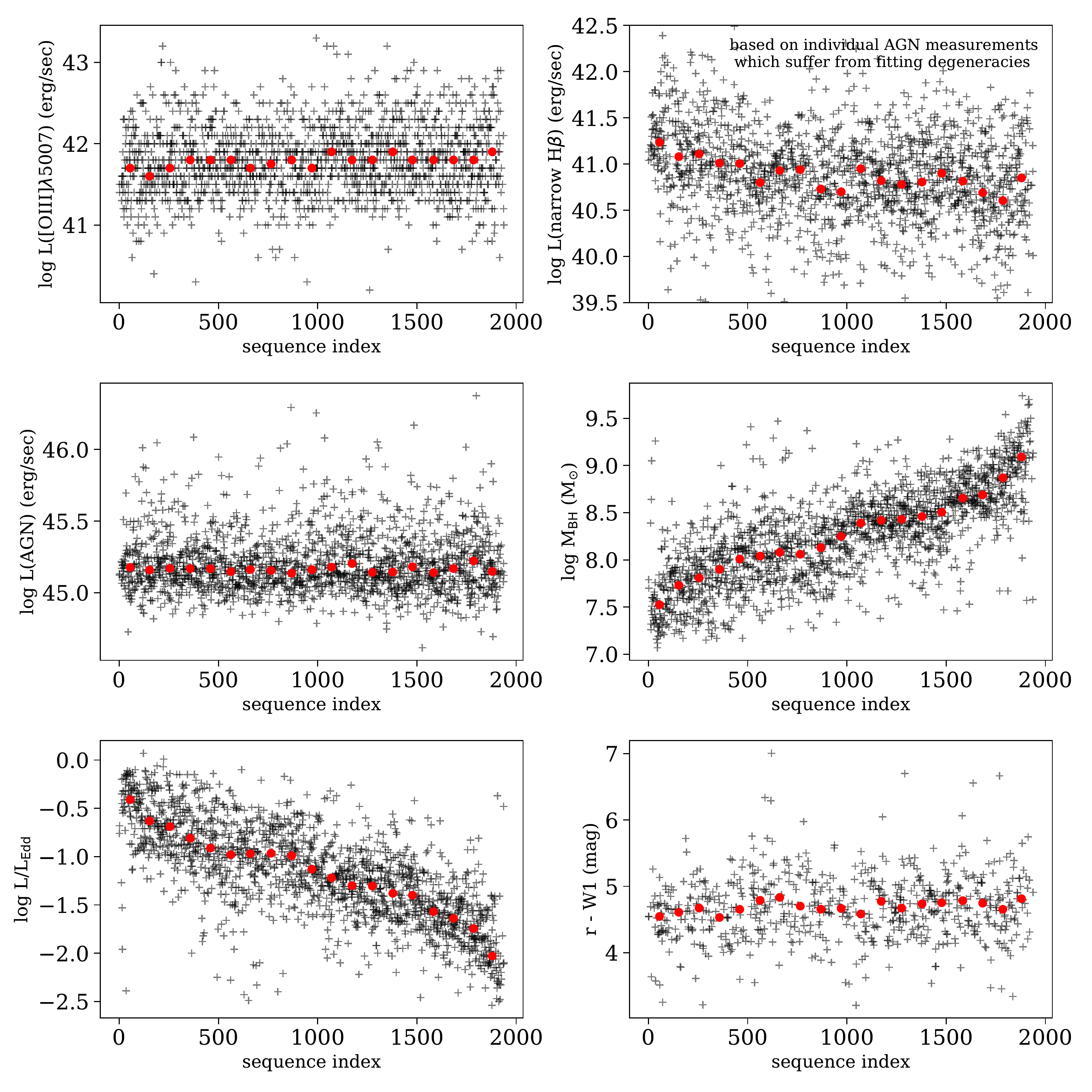}
\caption{Relation between the detected sequence and various measured AGN properties of individual sources, taken from \citet{shen11}. The grey crosses represent the narrow [OIII] luminosity, narrow H$\beta$ luminosity, AGN bolometric luminosity, black hole mass, Eddington ratio, and infrared colour (traced by $r-W1$), as a function of sequence index. The red points represent the median of these properties in sequence index bins.}
\label{f:correlation_with_sequence_index}
\end{figure*}

Figure \ref{f:correlation_with_sequence_index} shows the relation between the sequence detected in this work and different AGN properties. These properties are taken from the catalogue by \citet{shen11}, and are based on emission line and continuum fitting of individual sources. The sequence shows clear relations with the black hole mass and Eddington ratio, and with the narrow H$\beta$ line luminosity, albeit with a large scatter. We find negligible correlation between the sequence and the narrow [OIII] line luminosity, AGN bolometric luminosity, and infrared colour (traced by $r-W1$).

\subsection{Emission line decomposition}

\begin{figure*}
\includegraphics[width=0.9\textwidth]{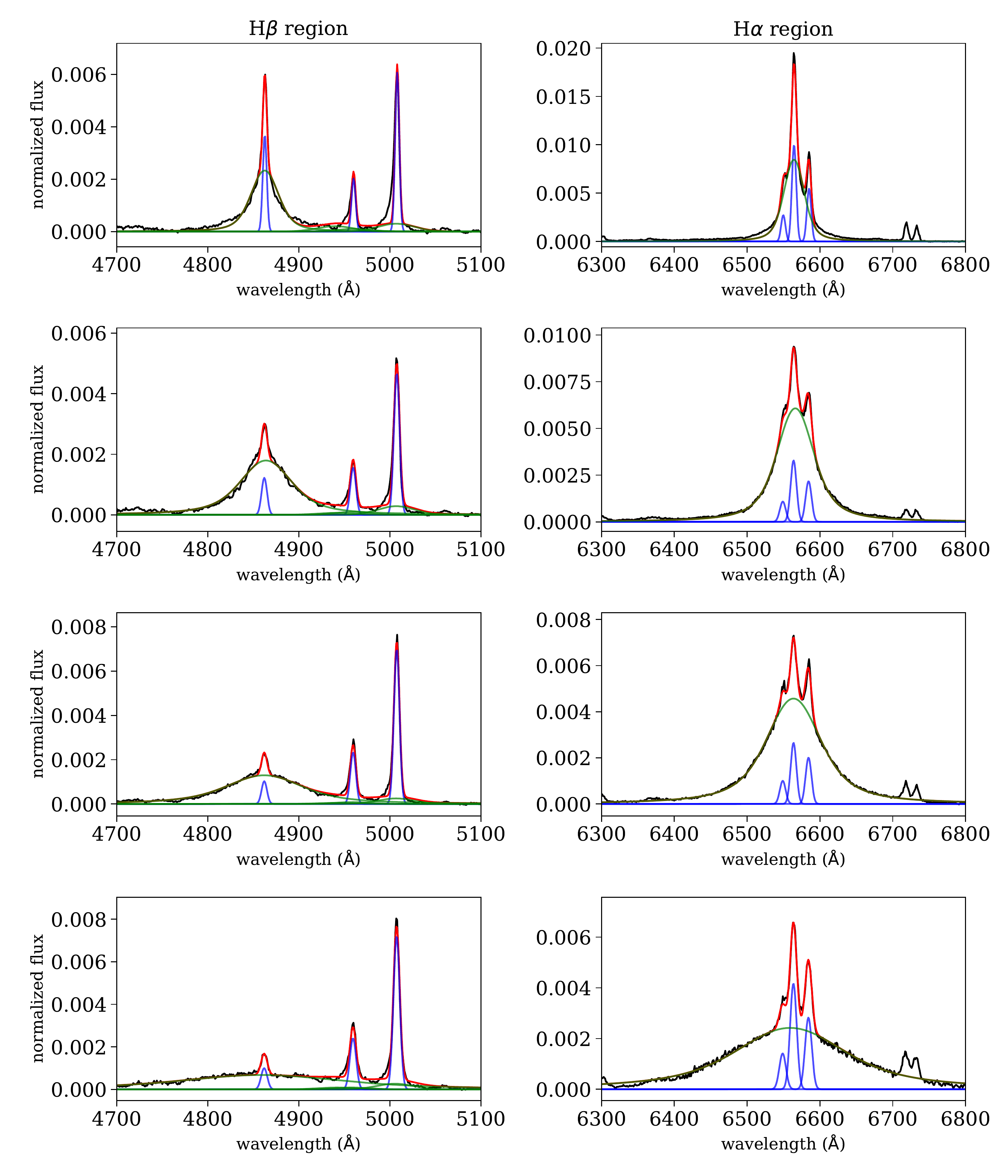}
\caption{Four examples of our best-fitting emission line profiles. Each row shows a median spectrum in the regions enclosing the H$\beta$+[OIII] lines (left) and H$\alpha$+[NII] lines (right). The median spectra are marked with black lines, the narrow lines are marked with blue, and the broad lines are marked with green. The full best-fitting profile is marked with red. See section \ref{s:corr_NLR_BLR} for additional details on the line decomposition. }
\label{f:stacked_spectra_fitting_example} 
\end{figure*}

Figure \ref{f:stacked_spectra_fitting_example} shows four examples of the emission line decomposition performed on the median type~I AGN spectra. The rows show median spectra, centred around the  H$\beta$+[OIII] lines (left panels) and  H$\alpha$+[NII] lines (right panels). The best-fit spectral energy distributions obtained with our parameterisation provide a satisfactory description of the data over the entire range of line shapes encountered with median spectra along the sequence.

\subsection{Contamination by star formation in the host galaxy}
\label{s:sf_contribution}

Recent IFU-based studies show that star forming galaxies that host AGN display SF-AGN mixing sequence through their optical line ratios, such that narrow emission lines, that include a contribution from \emph{both} young stars and AGN photoionisation, change gradually throughout the galaxy, and show line ratios that are consistent with AGN photoionisation close the central source, and with photoionisation by young stars farther away (e.g., \citealt{davies14}). A non negligible fraction of the host galaxies of the type~I AGN considered in this study are main sequence galaxies, and their narrow lines might include a contribution from photoionisation by young stars. It is therefore possible that the correlation we find between the narrow L([OIII])/L(H$\beta$) line ratio and FWHM(broad H$\alpha$) is driven by varying contribution of star formation in the host galaxy. 

To explore the possibility of star formation contamination in a systemic way, we focus on type~II AGN, for which the star formation rate can be estimated through the stellar continuum emission. We select type~II AGN from SDSS DR7, this time including systems that show contribution from both starburst and AGN activity, using the criterion by \citet{kauff03a}. We use the Dn4000-based star formation rate measurement from the MPA-JHU catalogue, and compute L(SF)/L(AGN) for each system in the sample. We then examine the narrow L([OIII])/L(H$\beta$) line ratio as a function of L(SF)/L(AGN). We find a correlation between L([OIII])/L(H$\beta$) and L(SF)/L(AGN) in systems with high L(SF)/L(AGN), such that L([OIII])/L(H$\beta$) decreases with increasing L(SF)/L(AGN). This suggests that the relative contribution of star formation versus the AGN drives the observed relation. However, the correlation is no longer detected for systems with log([OIII]/H$\beta$) $\gtrsim 0.55$, suggesting that star formation contamination is negligible in this regime. We therefore suggest that in this regime the correlation we find between L([OIII])/L(H$\beta$) and FWHM(broad H$\alpha$) is intrinsic to AGN gas properties.

\subsection{Fibre aperture effects}
     
The main result of this work, the scaling relation between the width of the broad H$\alpha$ line and the narrow line ratio L([OIII])/L(H$\beta$) has been obtained using fibre spectroscopy, i.e. measurements performed at a fixed angular scale. Sampling galaxies with a fixed fibre aperture impacts velocity dispersion estimates as a function of redshift. To explore this effect, we divided the type~II AGN sample into four redshift bins, and examined the best-fitting relation in Figure \ref{f:m_sigma_m_mbulge_final_version} and Eq.~\ref{eq:m-sigma-m-bulge} in the different bins. We find consistent best-fitting relations and scatters around them, suggesting that this fibre aperture effect is negligible in our case. Finally, we note that the resolution of the SDSS spectra is $69$ km/s per pixel. Figure~\ref{f:m_sigma_m_mbulge_final_version} shows a small fraction of objects for which the stellar velocity dispersion estimate is lower. These estimates are certainly more uncertain but we chose to show all available data points for completeness.

\section{Comparison to the PCA analysis from Boroson \& Green (1992)}
\label{ref:comparison_PCA}

\begin{figure}
\includegraphics[width=0.49\textwidth]{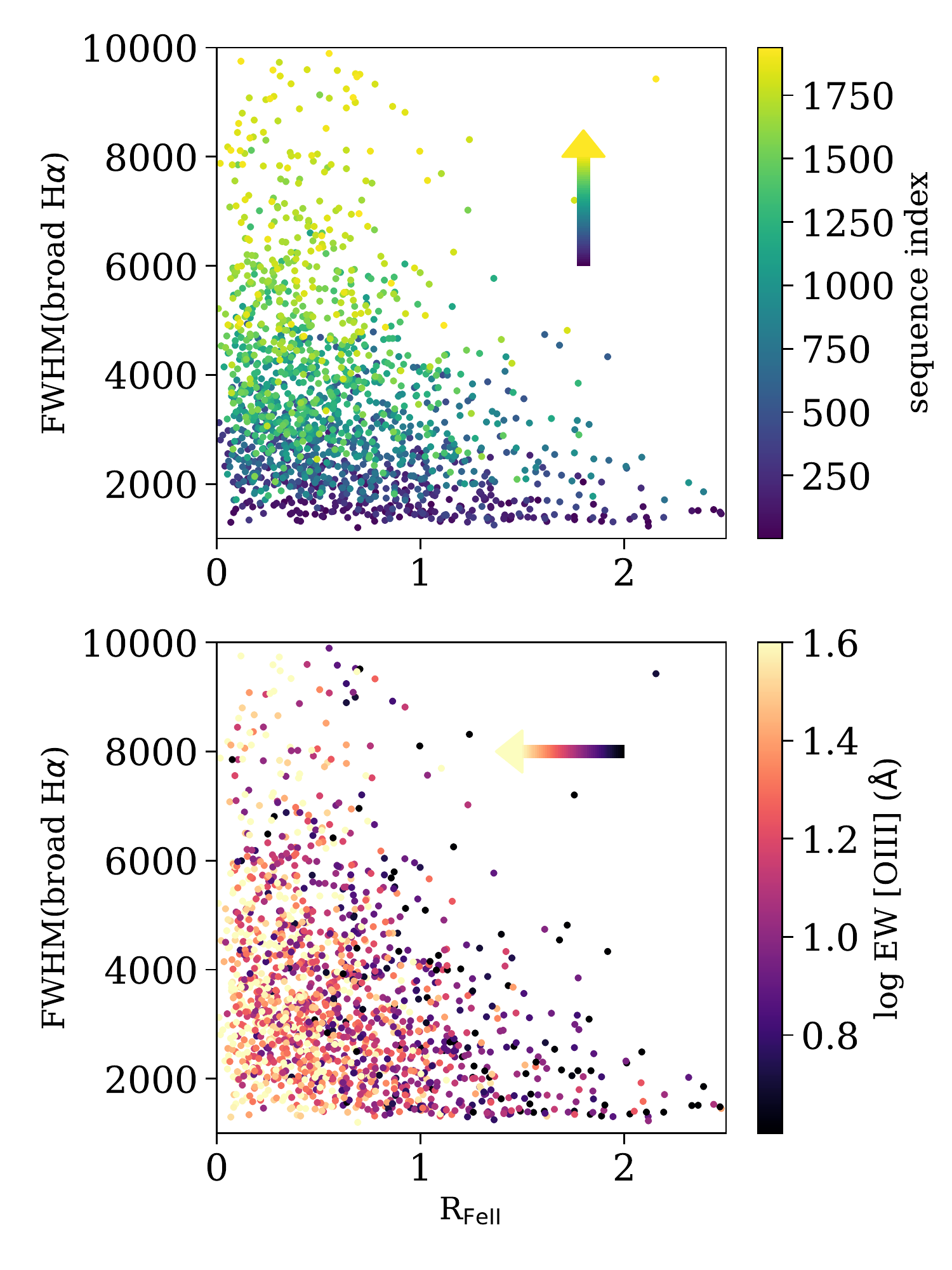}
\caption{The distribution of type~I AGN on the Eigenvector 1 plane as presented by \citet{shen14}. The FWHM(broad H$\alpha$) is plotted as a function of $R_{\mathrm{FeII}}$, and each measurement is colour-coded according to the sequence index (top panel) and [OIII] EW (bottom panel). While Eigenvector 1 is the horizontal trend in the FWHM(broad H$\alpha$)-$R_{\mathrm{FeII}}$ relation, the sequence in the vertical trend. For a given FWHM(broad H$\alpha$), the sequence index does not depend on $R_{\mathrm{FeII}}$.}
\label{f:relation_to_ev1}
\end{figure}

\citet{boroson92} pioneered dimensionality reduction studies of quasar spectra using principal component analysis. Their analysis did not use information from all the spectral pixels but was restricted to a specific set of parameters chosen by the authors to characterise certain properties of the continuum and emission lines. They analysed a sample of 87 broad line AGN and found that the main variance in the selected observables emerges from an anti-correlation between the equivalent width (EW) of the narrow [OIII] emission line and the strength of the broad FeII emission. More specifically, their Eigenvector 1 depends mainly on $R_{\mathrm{FeII}}$, defined as the ratio of Fe II EW to broad H$\beta$ EW, along which both [OIII] EW and the average FWHM(broad H$\beta$) decrease (see bottom panel of Figure \ref{f:relation_to_ev1}). This definition of Eigenvector 1 has been used by various authors over the past three decades. Recently, using type~I AGN observed by the SDSS, \citet{shen14} suggested that the Eddington ratio drives the Eigenvector 1 variation and that the scatter around it is due to orientation effects. 

It is natural to ask how the sequence presented in our work compares to the variations characterised by the generic Eigenvector 1. To stress the differences between the two, it is first worth pointing out that the so-called Eigenvector 1 is derived from the covariances of a set of measured parameters which does not include the luminosity of the narrow Balmer lines nor the narrow L([OIII])/L(H$\beta$) ratio used in our analysis. In addition, this PCA approach makes use of observables which are not included in our analysis, in particular the strength of FeII lines. Figure~\ref{f:relation_to_ev1} shows the distribution of the type~I AGN in our sample on the Eigenvector 1 plane, as presented by \citet{shen14}. The two panels show the FWHM(broad H$\alpha$) as a function of $R_{\mathrm{FeII}}$, colour-coded by our sequence index at the top and by the [OIII] EW at the bottom. When considering the distribution of points at fixed Balmer width, we observe a strong dependence on [OIII] EW as a function of $R_{\mathrm{FeII}}$. However, we do not see such a trend with our sequence index. In the upper panel, the variation in colours has a vertical gradient whereas the lower panel displays a horizontal one. This shows that the two trends or sequences are not aligned in the space of observables. While some of the Eigenvector 1 dependence overlaps with our sequence, mostly due to their mutual relation with the FWHM(broad H$\alpha$), our approach allows us to isolate one trend which is not explicitly visible through the classical PCA analysis of line properties. 

\begin{figure}
\includegraphics[width=0.49\textwidth]{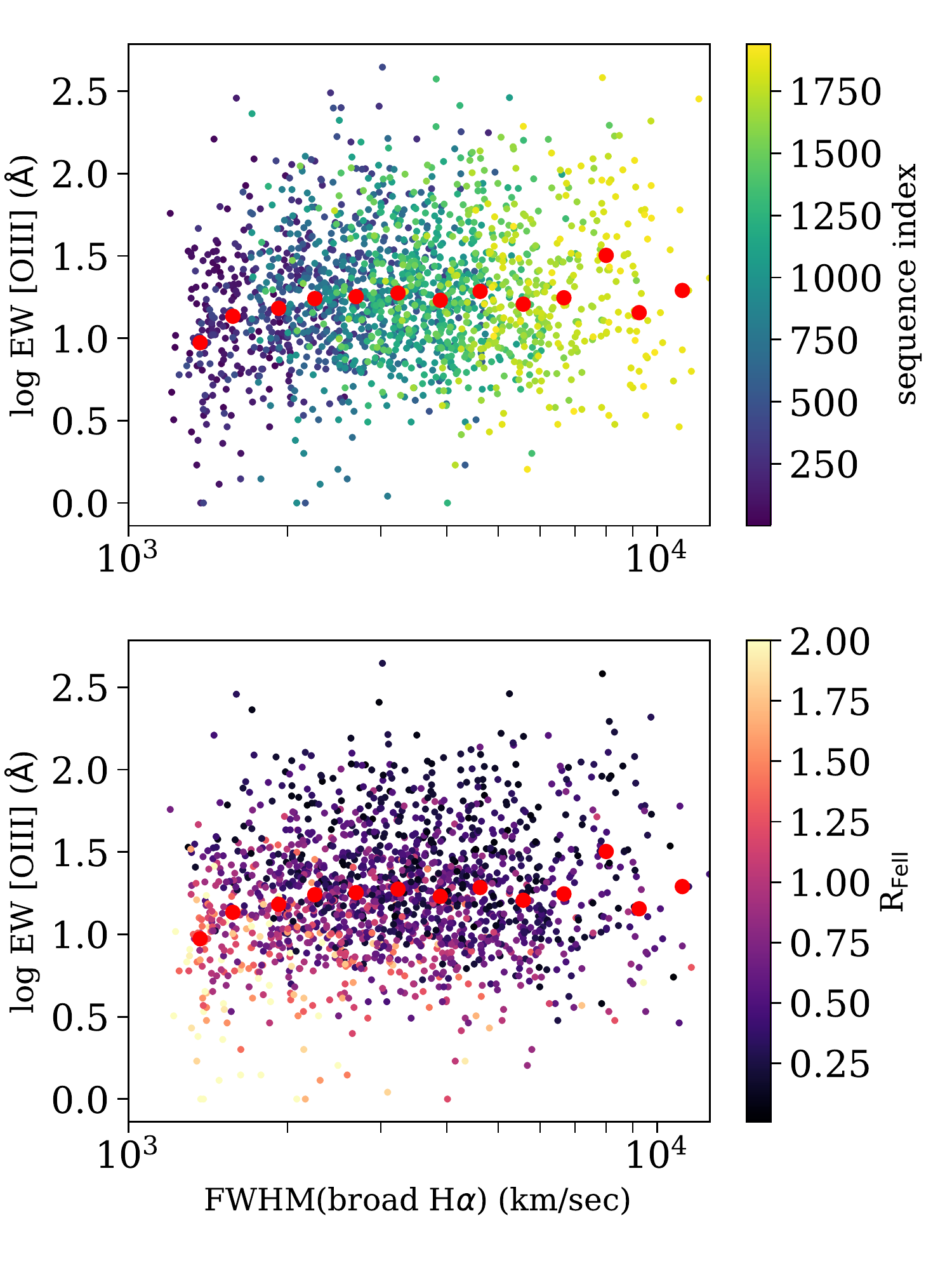}
\caption{The distributions of [OIII] EW as a function of FWHM(broad H$\alpha$), colour-coded by our sequence index (top panel) and by $R_{\mathrm{FeII}}$ (bottom panel), where the latter represents the traditional Eigenvector 1. The red points represent the median [OIII] EW as a function of FWHM(broad H$\alpha$). }
\label{f:relation_to_ev1_2}
\end{figure}

Another way to emphasise the difference between the two approaches is shown in Figure~\ref{f:relation_to_ev1_2}. In the [OIII] EW \& FWHM(broad H$\alpha$) plane, the standard Eigenvector 1 varies vertically whereas the gradient of our sequence index is in the horizontal direction. This figure also shows that [OIII] EW is not significantly correlated with FWHM(broad H$\alpha$). This suggests that the correlation we found between the narrow L([OIII])/L(H$\beta$) line ratio and the FWHM(broad H$\alpha$) is not a result of the [OIII] EW - FWHM(broad H$\alpha$) relation obtained through Eigenvector 1. Since the latter lacks the dynamical range required to produce the correlation we observe (over one order of magnitude in L([OIII])/L(H$\beta$) line ratio), our result cannot be reproduced through the Eigenvector 1. 

So, while both the standard PCA sequence and the one introduced in this work correlate with FWHM(broad H$\beta$), they do not trace the same correlations nor physical effects. Here it is important to note that the decrease in FWHM(broad H$\beta$) originates from lack of objects with high FWHM(broad H$\beta$) and high $R_{\mathrm{FeII}}$ (\citealt{shen14} attributed this to an orientation effect). At fixed FWHM(broad H$\beta$), our sequence has no knowledge of $R_{\mathrm{FeII}}$ nor [OIII] EW. The two approaches are describing different phenomena.

\end{document}